\documentclass[twocolumn]{aastex631}

\usepackage{natbib}
\usepackage{amssymb}
\usepackage{amsmath}
\usepackage{verbatim}
\usepackage{indentfirst}
\usepackage{xcolor}
\usepackage{graphicx}
\usepackage{hyperref}
\usepackage{makecell}

\newcommand{\Rnum}[1]{\uppercase\expandafter{\romannumeral #1\relax}}

\newcommand{\NII}{[\ion{N}{2}]}

\newcommand{\SII}{[\ion{S}{2}]}

\newcommand{\ha}{{\rm H\ensuremath{\alpha}}}
\newcommand{\hb}{{\rm H\ensuremath{\beta}}}

\newcommand{\mbh}{{\ensuremath{M\raisebox{-.3ex}{$\bullet$}}}}
\newcommand{\Mbh}{{\ensuremath{\rm M\raisebox{-.3ex}{$\bullet$}}}}

\newcommand{\msun}{{\ensuremath{{\rm M_\odot}}}}

\newcommand{\valerrud}[3]{${#1}^{+#2}_{-#3}$}

\newcommand{\angstrom}{\textup{\AA}}

\newcommand{\jav}{{\rm Javelin}}
\newcommand{\zdcf}{{$z$DCF}}

\newcommand{\snu}{\affil{Department of Physics \& Astronomy, Seoul National University, Seoul 08826, Republic of Korea}}
\newcommand{\kasi}{\affil{Korea Astronomy and Space Science Institute, Daejeon 34055, Republic of Korea}}
\newcommand{\umich}{\affil{Department of Astronomy, University of Michigan, Ann Arbor, MI 48109, USA}}
\newcommand{\nasa}{\affil{NASA/GSFC, Code 662, Greenbelt, MD 20771, USA}}
\newcommand{\nysc}{\affil{National Youth Space Center, Goheung 59567, Republic of Korea}}
\newcommand{\ucla}{\affil{Department of Physics and Astronomy, University of California, Los Angeles, CA 90095-1547, USA}}
\newcommand{\calpoly}{\affil{Physics Department, California Polytechnic State University, San Luis Obispo, CA 93407, USA}}
\newcommand{\yonsei}{\affil{Department of Astronomy, Yonsei University, Seoul 03722, Republic of Korea}}
\newcommand{\knu}{\affil{Major in Astronomy and Atmospheric Sciences, Kyungpook National University, Daegu 41566, Republic of Korea}}
\newcommand{\uci}{\affil{Department of Physics and Astronomy, 4129 Frederick Reines Hall, University of California, Irvine, CA, 92697-4575, USA}}
\newcommand{\cbu}{\affil{Department of Astronomy and Space Science, Chungbuk National University, Cheongju 28644, Republic of Korea}}
\newcommand{\spacebeam}{\affil{Spacebeam Inc., Cheongju 28165, Republic of Korea}}

\begin{document}

\title{The Seoul National University AGN Monitoring Project \Rnum{4}: H$\alpha$ reverberation mapping of 6 AGNs and the H$\alpha$ Size-Luminosity Relation}

\author[0000-0003-2010-8521]{Hojin Cho} \snu
\author[0000-0002-8055-5465]{Jong-Hak Woo} \snu
\author[0000-0002-2052-6400]{Shu Wang} \snu
\author[0000-0002-4704-3230]{Donghoon Son} \snu
\author[0000-0001-6363-8069]{Jaejin Shin} \snu \knu \kasi
\author[0000-0002-8377-9667]{Suvendu Rakshit} \snu \affiliation{Aryabhatta Research Institute of Observational Sciences, Manora Peak, Nainital-263001, Uttarakhand, India}

\author[0000-0002-3026-0562]{Aaron J.\ Barth} \uci
\author[0000-0003-2064-0518]{Vardha N. Bennert} \calpoly
\author[0000-0001-5802-6041]{Elena Gallo} \umich
\author[0000-0002-2397-206X]{Edmund Hodges-Kluck} \nasa \umich
\author[0000-0002-8460-0390]{Tommaso Treu} \ucla

\author[0000-0001-5134-5517]{Hyun-Jin Bae} \snu \yonsei
\author[0000-0002-4896-770X]{Wanjin Cho} \snu
\author{Adi Foord} \affil{Kavli Institute of Particle Astrophysics and Cosmology, Stanford University, Stanford, CA 94305, USA}
\author{Jaehyuk Geum} \knu
\author{Yashashree Jadhav} \snu
\author{Yiseul Jeon} \snu
\author[0000-0003-2632-8875]{Kyle M. Kabasares} \uci
\author{Daeun Kang} \snu
\author[0000-0003-3390-1924]{Wonseok Kang} \nysc \spacebeam
\author[0000-0002-2156-4994]{Changseok Kim} \snu
\author{Donghwa Kim} \snu \affil{Graduate School of Data Science, Seoul National University, Seoul 08826, Republic of Korea}
\author[0000-0002-3560-0781]{Minjin Kim} \knu
\author[0000-0003-4686-5109]{Taewoo Kim} \nysc \cbu \spacebeam
\author[0000-0003-1270-9802]{Huynh Anh N. Le} \snu \affiliation{CAS Key Laboratory for Research in Galaxies and Cosmology, Department of Astronomy, University of Science and Technology of China, Hefei 230026, China}
\author[0000-0001-6919-1237]{Matthew A. Malkan} \ucla
\author{Amit Kumar Mandal} \snu
\author{Daeseong Park} \kasi \knu
\author{Songyoun Park} \snu
\author[0000-0001-9515-3584]{Hyun-il Sung} \kasi
\author[0000-0002-1912-0024]{Vivian U} \uci
\author[0000-0002-4645-6578]{Peter R. Williams} \ucla

\correspondingauthor{Jong-Hak Woo}\email{woo@astro.snu.ac.kr}


\begin{abstract}
The broad line region (BLR) size-luminosity relation has paramount importance for estimating the mass of black holes in active galactic nuclei (AGNs). Traditionally, the size of the H$\beta$ BLR is often estimated from the optical continuum luminosity at 5100\angstrom{} , while the size of the H$\alpha$ BLR and its correlation with the luminosity is much less constrained. As a part of the Seoul National University AGN Monitoring Project (SAMP) which provides six-year photometric and spectroscopic monitoring data, we present our measurements of the H$\alpha$ lags of 6 high-luminosity AGNs. Combined with the measurements for 42 AGNs from the literature, we derive the size-luminosity relations of H$\alpha$ BLR against broad H$\alpha$ and 5100\angstrom{} continuum luminosities. We find the slope of the relations to be $0.61\pm0.04$ and $0.59\pm0.04$, respectively, which are consistent with the \hb{} size-luminosity relation. Moreover, we find a linear relation between the 5100\angstrom{} continuum luminosity and the broad H$\alpha$ luminosity across 7 orders of magnitude. Using these results, we propose a new virial mass estimator based on the H$\alpha$ broad emission line, finding that the previous mass estimates based on the scaling relations in the literature are overestimated by up to 0.7 dex at masses lower than $10^7$~M$_{\odot}$.
\end{abstract}


\section{Introduction}
Mass is the most important physical property of a black hole that we can measure. While the mass of a black hole can be determined by measuring its gravitational radius, this method thus far has been applied to only two black holes \citep{EHT_M87_VI, EHT_Sgr_IV}. For most extragalactic black holes, the mass is instead measured by observing the kinematics of orbiting bodies near the black hole. While the mass of several black holes has been measured via the kinematics of surrounding gas \citep[e.g.,][]{Scharwaechter+13, denBrok+15, GRAVITY+18Nat, Kabasares+22} or stars \citep[e.g.,][]{vanderMarel+94, Nguyen+18, Nguyen+19}, this technique requires an exceptional angular resolution that can resolve the sphere of influence of the black hole \citep{Peebles72}.

The mass of active galactic nuclei (AGNs) can be measured without good spatial resolution via reverberation mapping \citep{Blandford&McKee82}. By measuring the time delay ($\tau$) of the broad emission line flux against the continuum, combined with its line width ($\varDelta V$) measured from the spectrum, the mass of the black hole (\mbh) can be determined using
\begin{equation}
\begin{aligned}
\Mbh = f \frac{c\,\tau\cdot\varDelta V^2}{G} \label{eq:mbh}
\end{aligned}
\end{equation}
where $c$ is the vacuum speed of light and $G$ is the gravitational constant. The virial factor, $f$, is a dimensionless scale factor reflecting the geometry and kinematics of the BLR. To date, the masses of more than 100 AGNs have been measured by applying this technique to broad \hb{} lines \citep[e.g.,][]{Bentz&Katz15}.

This method can be extended to a far larger number of black holes using the size-luminosity relation of the \hb{}-emitting zone of the BLR by estimating the time lag from the 5100\angstrom{} luminosity of the AGN accretion disk \citep[e.g.,][]{Kaspi+00, Bentz+13}. It provides a short-cut to estimating the broad line time lag without going through a reverberation mapping campaign, enabling a way to estimate the mass of the black hole with a single spectroscopic observation; hence, it is called the single-epoch method.

Compared to \hb{}, the \ha{}-emitting zone of BLR has been relatively unexplored, currently with the \ha{} lag measurements of only $\sim$50 AGNs \citep[e.g.,][]{Kaspi+00, Bentz+10, Grier+17}. This is because observing \ha{} poses more challenges than observing \hb{}. For instance, \ha{} is in the redder part of the optical spectrum, making it vulnerable to the airglow lines as well as Fraunhofer A and B band telluric absorption lines, given appropriate redshifts. Moreover, there is no strong narrow line in the vicinity of \ha{} that could be used for flux calibration, whereas the calibration of \hb{} emission lines can utilize the invariant and strong fluxes of [\ion{O}{3}] narrow lines.

Nevertheless, the benefit of using broad \ha{} lines for the single-epoch method outweighs its difficulty. First, measuring \ha{} flux is more reliable than measuring \hb{}. The \ha{} line is stronger than the \hb{} line by at least a factor of 3, and this factor increases to 4-6 for broad emission lines \citep{Netzer90}. Some AGNs even exhibit a relatively weak, if present at all, broad \hb{} emission \citep{Osterbrock81}.

Furthermore, broad line fluxes of \ha{}, as well as \hb{}, can be measured with less degeneracy than the continuum luminosity, making it an ideal proxy for the size-luminosity relation. The observed continuum luminosity in the AGN spectrum is contaminated by the starlight from its host galaxy or synchrotron radiation from the jet in the case of radio-loud AGNs, which must be removed to use the relation. The removal of host stellar emission can be achieved either by modeling the image of the AGN to determine the host galaxy fraction to the AGN spectrum with high-resolution images \citep[e.g.,][]{Bentz+13} or by decomposing the continuum spectrum as a sum of the stellar and AGN components in the high S/N spectra \citep[e.g.,][]{Park+12}. The removal of jet contamination would require multi-wavelength observations \citep[e.g.,][]{Paltani+1998, Soldi+2008}. The broad \ha{}/\hb{} lines, on the other hand, are purely from the BLR of the AGN and can be separated from narrow lines.

There is, however, one difficulty in using \ha{} for single-epoch mass estimation: a size-luminosity relation involving \ha{} luminosity has not yet been reported. As a workaround, \citet{Greene&Ho05} demonstrated the empirical relation between the broad \ha{} line luminosity and the 5100\angstrom{} continuum luminosity and proposed to use it in conjunction with the \hb{} size-luminosity relation to construct a \ha{}-based single-epoch mass estimator. To date, it has been applied to a number of AGNs that are too faint to measure the AGN 5100\angstrom{} luminosity and/or \hb{} line width correctly. In particular, the masses of low-luminosity AGNs and active intermediate-mass black holes (IMBHs) were measured using this recipe \citep[e.g.,][]{Reines+13, Shin+22}, which suffers from substantial uncertainty due to the scatter of the scaling relations. Therefore, a relation between the size of the \ha{} BLR and the broad \ha{} luminosity will provide more robust estimations.

The Seoul National University AGN Monitoring Project \citep[SAMP;][]{Woo+19b, Rakshit+19} is a reverberation mapping campaign aimed at the \hb{} time lags of dozens of high-luminosity AGNs to expand the size-luminosity relation toward a higher luminosity regime. In this paper, we present the SAMP results on \ha{} time lag measurements and demonstrate the new empirical relation between the \ha{} BLR size and the broad \ha{} luminosity. In section~\ref{s:obsred}, we describe the data acquisition and reduction. In section~\ref{s:specanal}, we perform spectral decomposition and \ha{} flux measurements. The time lag measurements are provided in section~\ref{s:timelag}. Section~\ref{s:sl} presents the size-luminosity relation of the \ha{} broad line. We discuss the implications of this size-luminosity relation in section~\ref{s:discuss}. Section~\ref{s:concl} gives a brief summary of the paper.
Throughout this paper, we adopt a flat $\Lambda$CDM cosmology with $H_0 = 72\,{\rm km\, s^{-1}\, Mpc^{-1}}$ and $\Omega_{\rm m} = 0.3$.


\section{Observations and Data Reduction}\label{s:obsred}
The initial sample of SAMP observations consisted of 100 AGNs selected from the literature, which was described in detail by \citet{Woo+19b}. To briefly summarize, 85 AGNs in local universe ($z<0.5$) with $V<17$ were selected from the Million Quasars catalog \citep[MILLIQUAS,][]{HMQ, Milliquas}, whose observed-frame lags were expected to be $40<(1+z)\tau_\mathrm{H\beta}<250$ days based on the R-L relation of \citet{Bentz+13}. The other 15 AGNs were selected from the Palomar-Green catalog \citep{Boroson&Green92}. During the first few years, we were able to identify AGNs with very low variability. Note that since the expected lag of the sample is relatively large, we were able to predict whether the line flux would vary at each epoch based on the photometric light curves. By selecting the most variable sources, we narrowed down the sample to 32 objects for continuous monitoring for six years, by excluding objects with weak variability. In this paper, we specifically focus on 13 objects of which \ha{} lines were observable with our spectral configurations.

\subsection{Photometry}
We carried out our photometric monitoring observations using several telescopes, including MDM 1.3 m and 2.4 m telescopes, the Lemmonsan Optical Astronomy Observatory (LOAO) 1 m telescope, the Lick observatory 1 m nickel telescope, the Las Cumbres Observatories Global Telescope (LCOGT) network, and the Deokheung Optical Astronomy Observatory (DOAO) 1 m telescope. The acquisition of the photometric images, reduction processes, and photometry are described in our previous paper \citep{samp_hb_inprep}. Here, we used fully reduced and intercalibrated B and V band light curves. Typically, the B-band light curves spanned $\sim$2000 days with a median cadence of 4 days, resulting in $\sim$250 epochs, except for Mrk~1501, which was observed for 115 epochs over 1740 days with a median cadence of 6 days. The V-band light curves were acquired with a median cadence of one week.

\subsection{Spectroscopy}\label{ss:specred}
Spectroscopic observations of \ha{} lines were carried out using the Shane 3 m telescope, located at the Lick observatory on Mt. Hamilton, California, USA. Note that while we used the Lick 3 m and MDM 2.4 m telescopes for the SAMP, we only use the Lick 3 m data, which covers the \ha{} line. The details of the spectroscopic observations of the SAMP were described by \citet{Rakshit+19}.

We used the Kast double spectrograph,\footnote{\url{https://mthamilton.ucolick.org/techdocs/instruments/kast/}} which employs dichroic beamsplitters to acquire the red side and blue side spectra simultaneously. We used the red-side spectra with a 600 lines/mm grating. At the beginning of the campaign, the wavelength coverage of our spectra was 4450-7280~\angstrom{} with 2.33\angstrom{} /pixel sampling, and the spectra obtained during this period are hereafter denoted as \emph{early configuration spectra}. In September 2016, the detector was replaced with a 2K$\times$4K CCD, covering 4750-8120~\angstrom{} with 1.27\angstrom{} /pixel sampling, and the wavelength coverage was slightly adjusted to 5050-8424~\angstrom{} in March 2019. Spectra obtained after September 2016 are hereafter denoted as \emph{late configuration spectra}, which constitutes 80\% of epochs. We used a 4\arcsec{} slit width to minimize the slit loss. The instrumental resolving power is $R=650$, which was measured from unblended airglow lines near 7500\angstrom{} . This corresponds to the FWHM velocity of $\rm 460\, km\,s^{-1}$. Note that the actual resolution of AGN spectra would be better than what is measured from the night sky emissions since the slit width is wider than the seeing FWHM (1\farcs{}5-4\arcsec{}).

Each night, we obtained the bias, arc, and flat frames at the beginning and end of the night. Note that the arc lamp images were taken using the 0\farcs{}5 width slit to improve the accuracy of the wavelength solution. We also observed at least one of the spectrophotometric standard stars listed by \citet{Oke90}, and any spectra taken on nights without spectrophotometric stars were discarded from the \ha{} analysis.

The red-side spectra of Lick/Kast were preprocessed primarily using \texttt{PypeIt} v1.4\citep{PypeIt:JOSS, PypeIt:Zenedo}. This pipeline was chosen to minimize the human intervention in the fitting of the wavelength solution and the sensitivity function. The latter is particularly susceptible to human factors due to the highly variable telluric OH absorption band near the red-side edge of the spectra. We created pixel flats and traced the slit using dome flat-field images. The wavelength solutions were derived using Ne and Ar lines in the arc frames in full template mode, and barycentric corrections were applied to each object frame. We used the optimal extraction algorithm \citep{Horne86}, implemented in \texttt{PypeIt}, to obtain photon-count spectra because optimally extracted spectra yield higher S/N than those produced with standard aperture extraction. The optimal extraction algorithm is generally not recommended for extended objects such as AGNs with resolved NLRs \citep[e.g.,][]{Barth+15}. However, we confirm that all objects in our campaign did not show extended narrow lines, even under the best seeing conditions (typically $\leq$1\farcs{}5). Furthermore, the resulting optimally extracted spectra showed no differences compared to the aperture-extracted spectra, except for having a higher S/N.

\subsubsection{Flux calibration}
We visually inspected all spectra of spectrophotometric standard stars. After masking strong Balmer absorption lines, sensitivity functions were constructed from each reliable standard star spectrum by jointly fitting it with polynomial functions of wavelength and the model of telluric absorption lines using a script provided by \texttt{PypeIt} in IR mode. This yielded 1-6 different sensitivity functions per successful night. We derived the median value of the individual sensitivity function, and the function that showed a median sensitivity of the given night was chosen to be the representative sensitivity function of that night.

Spectra of AGNs and standard stars were then calibrated using the representative sensitivity for that night. We discarded any spectra taken on the nights that failed to produce at least one reliable sensitivity. Then, the atmospheric extinction was corrected based on the airmass difference between the sensitivity function and the object frame.

To further calibrate the flux in each spectrum, we compared the synthetic V-band flux obtained from the spectrum with the photometric light curves. First, we constructed the \jav{} model \citep{Zu+16} of the photometric V-band light curve and interpolated it onto each spectral epoch. The B-band light curve, which has a much shorter cadence, was jointly modeled with the V-band to improve the quality of the interpolation. Then, each spectrum, after being multiplied by the V-band filter transmission curve, was integrated to synthesize the V-band flux. Finally, we scaled the spectra so that the synthetic V-band flux was the same as the photometric flux. Note that the \emph{late configuration spectra} did not cover the entirety of the V-band bandwidth. For these epochs, we calculated the portion of V-band flux that was included in each spectrum based on \emph{early configuration spectra}, which was $\sim$80\% on average, assuming that the spectral shape of AGNs did not change significantly over the campaign period.

\subsubsection{Telluric correction}
\begin{figure*}[!thpb]
\centering
\includegraphics[width=0.95\textwidth]{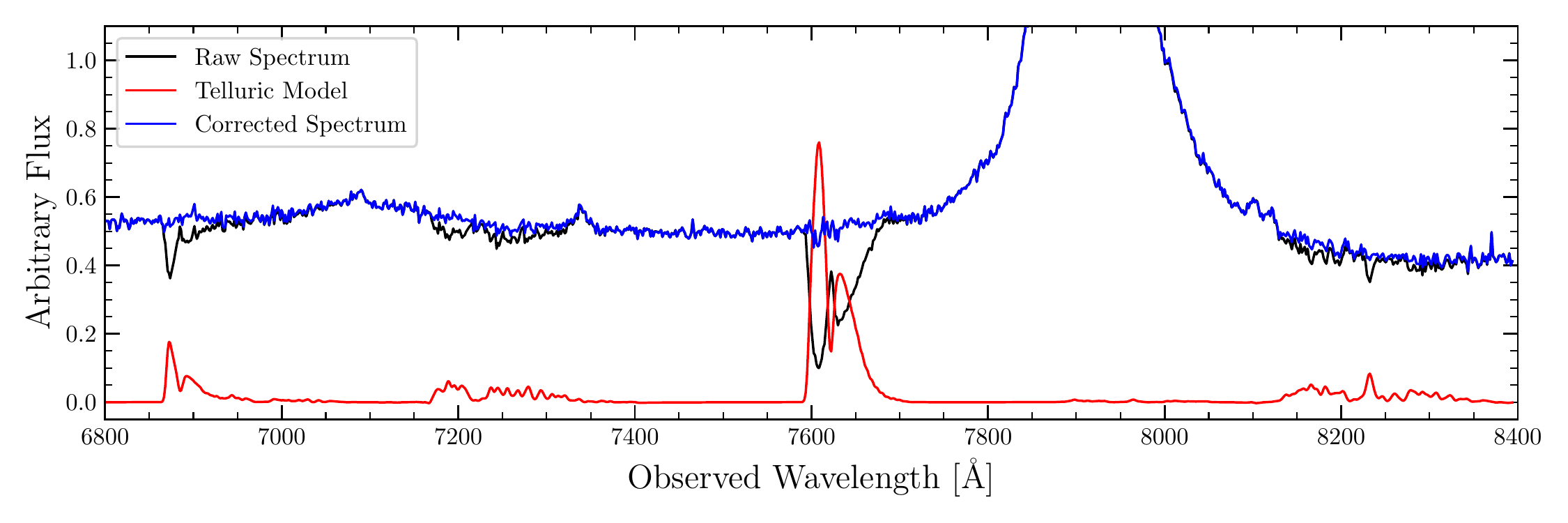}
\caption{An example of telluric correction. The black line represents the flux-calibrated spectrum of PG~0947$+$396 without the telluric correction. Red represents the telluric model in the optical depth unit (arbitrarily scaled), with the velocity and velocity dispersion adjusted to fit the AGN spectrum. Blue is the AGN spectrum after correcting for telluric absorption. \label{fig:tellcor}}
\end{figure*}

The telluric absorption features were corrected using a set of atmospheric absorption line models of the Lick observatory provided by \texttt{PypeIt}. This consists of 28,413 high-spectral resolution model spectra of telluric absorption lines. We first performed principal component analysis (PCA) on the models and chose the largest 25 components. Then, we fitted the calibrated spectra of spectrophotometric standards with PCA components using \texttt{pPXF} \citep{pPXF} to obtain the weight for each component to reconstruct a high-resolution telluric model for each standard spectrum. We averaged the telluric models for each night to create the nightly telluric model spectrum. Then, for each object spectrum, we masked the narrow lines and fitted its red part ($\lambda_\mathrm{obs} > 6500\,\mathrm{\angstrom{}}$ ) with the nightly model and polynomial functions using \texttt{pPXF}. After shifting and broadening the model by the best-fit parameters provided by \texttt{pPXF}, we divided the spectrum by the model to generate a telluric corrected spectrum of our targets. An example of telluric correction is shown in Figure~\ref{fig:tellcor}. After correcting for the telluric absorption lines, the corrected flux level is consistent with the nearby continuum. However, we note strong residuals for a couple of spectra for each object after removing the telluric lines, which is presumably due to the decreased signal-to-noise ratio. Nevertheless, these residual features did not affect our spectral analysis, as the residual features are far narrower than the AGN emission lines.

\subsubsection{Shift correction}
Despite the wavelength calibration and the barycentric corrections, the wavelength solutions deviate between the different spectra by several angstroms because of instrumental flexure and/or pointing accuracy within the slit. To compensate for this, we shifted each spectrum so that the peak of the \ha{} line falls exactly on the theoretical wavelength of \ha{}. Note that the \ha{} line is easier to use for this purpose compared to much weaker narrow emission lines. We first calculated the derivatives of the individual spectra by applying a Savitzky-Golay filter using \texttt{SciPy} \citep{SciPy} with a window width of $\sim$ 1260 $\rm km\,s^{-1}$, and the peak of \ha{} was calculated by finding the root of the derivatives. The window width was chosen based on experiments so that the secondary peaks due to \NII{} lines were smoothed out. Finally, we shifted and resampled the spectra.

Among the full sample of 32 AGNs that were monitored, 13 objects showed \ha{} in the observed spectra, depending on the redshift. Four of them suffered from strong telluric absorption because the Fraunhofer A-band fell on the very center of the \ha{} line. For these objects, the described telluric correction was unreliable, so we did not analyze it further. Additionally, there were fewer than 20 spectra available for 3 objects, which is unsuitable for time-series analysis. We discarded these objects as well. We present the analysis of the remaining 6 objects, whose properties are summarized in Table \ref{table:objects}.

\renewcommand{\arraystretch}{1.25}
\begin{deluxetable*}{rlcccccccccl}
\tablewidth{0.99\textwidth}
	\tablecolumns{12}
	\tablecaption{\ha{} Objects\label{table:objects}}
	\tablehead
	{
		& \colhead{Name} &\colhead{SDSS Identifier}& \colhead{R.A.} & \colhead{Dec.}& \colhead{$z$} & \colhead{$A_{V}$} & \colhead{$N_\mathrm{ph}$}& \colhead{$\Delta t_\mathrm{ph}$}& \colhead{$N_\mathrm{sp}$}& \colhead{$\Delta t_\mathrm{sp}$} & \colhead{SAMPID}
		\\
		& \colhead{(1)}    &\colhead{(2)}       &\colhead{(3)}&\colhead{(4)}&\colhead{(5)}&\colhead{(6)}&\colhead{(7)}&\colhead{(8)}&\colhead{(9)}&\colhead{(10)}&\colhead{(11)}
	}
	\startdata
1. & Mrk~1501      & J001031.00$+$105829.4 & 00:10:31.0 & +10:58:29.5 & 0.0893 & 0.269 & 115 & 6 & 22 & 37 & P02 \\
2. & J0101$+$422   & J010131.17$+$422935.5 & 01:01:31.1 & +42:29:36.0 & 0.1900 & 0.238 & 255 & 4 & 38 & 22 & Pr1\_ID01 \\
3. & PG~0947$+$396 & J095048.39$+$392650.4 & 09:50:48.4 & +39:26:50.5 & 0.2059 & 0.052 & 267 & 4 & 31 & 29 & Pr1\_ID15 \\
4. & J1217$+$333   & J121752.16$+$333447.2 & 12:17:52.2 & +33:34:47.3 & 0.1784 & 0.036 & 246 & 4 & 25 & 31 & Pr1\_ID29 \\
5. & VIII~Zw~218   & J125337.71$+$212618.2 & 12:53:37.7 & +21:26:18.2 & 0.1274 & 0.135 & 239 & 4 & 38 & 29 & Pr1\_ID30 \\
6. & PG~1440$+$356 & J144207.47$+$352622.9 & 14:42:07.5 & +35:26:23.0 & 0.0791 & 0.038 & 235 & 4 & 34 & 29 & Pr2\_ID26 \\
	\enddata
\tablecomments{Columns are (1) object name, (2) the SDSS identifier, (3) right ascension (J2000), (4) declination (J2000), (5) redshift (from \citeauthor{ned}), (6) galactic extinction in V-band by \citet{SandF11}, (7) number of epochs in photometric light curve, (8) median cadence of photometric light curve, (9) number of epochs in spectroscopic light curve, (10) median cadence of spectroscopic light curve, and (11) SAMPID (refer to \citealt{Woo+19b}).
    }
\end{deluxetable*}


\section{Spectral Analysis}\label{s:specanal}

Many of the early reverberation mapping studies measured the broad \ha{} line flux by directly integrating the spectrum within a fixed range without fitting the line profile with a model. In this case, the continuum below the emission line was fitted with a straight line at the two ends of the \ha{} line profile \citep[e.g.,][]{Kaspi+00, Bentz+10}. While this procedure is straightforward, we found it insufficient for our objects at higher redshift. First, the \ha{} emission line of our objects is located close to the edge of the detector, making it challenging to directly determine the continuum level. This is further complicated by the presence of telluric OH absorption lines near the edge of the detector. Although the absorption was averaged out through correction, some epochs exhibited strong residuals due to velocity mismatch. Finally, some of our AGNs displayed moderate Fe II lines in the blue-side continuum of \ha{}. While their fluxes are relatively small compared to the continuum or \ha{} emission, they are still strong enough to influence the slope of the continuum fit, thereby reducing the accuracy of the \ha{} flux. This issue is similar to what was pointed out by \citet{Barth+15} regarding the construction of \hb{} light curves. It is therefore preferable to model the line profiles and measure the broad \ha{} flux at each epoch.

To do this, we first constructed the mean spectrum for each object. We modeled its continuum as a power law along with the \ion{Fe}{2} lines based on the model by \citet{Boroson&Green92} using suitable windows, i.e., 4175-4250~\angstrom{} , 4500-4725~\angstrom{} , 5090-5780~\angstrom{} , 6000-6280~\angstrom{} , and 6800-7650~\angstrom{} , in the rest frame, if covered by the spectrograph. The portions of the spectrum that showed either (1) strong telluric residual or (2) strong narrow lines were masked before the continuum fitting, leaving at least one window on the blue side of \hb{}, one between \ha{} and \hb{}, and one in the red side of \ha{}. We did not include the stellar host continuum in the model since our 6 objects do not exhibit strong stellar absorption features. We determined the best-fit model based on the maximum-likelihood method using the \texttt{zeus} MCMC sampler \citep{zeus:method, zeus:program}. The best-fit models of the continuum and \ion{Fe}{2} lines were subtracted from the mean spectrum, leaving the line spectrum only.

\begin{figure*}[!thpb]
\centering
\includegraphics[width=0.95\textwidth]{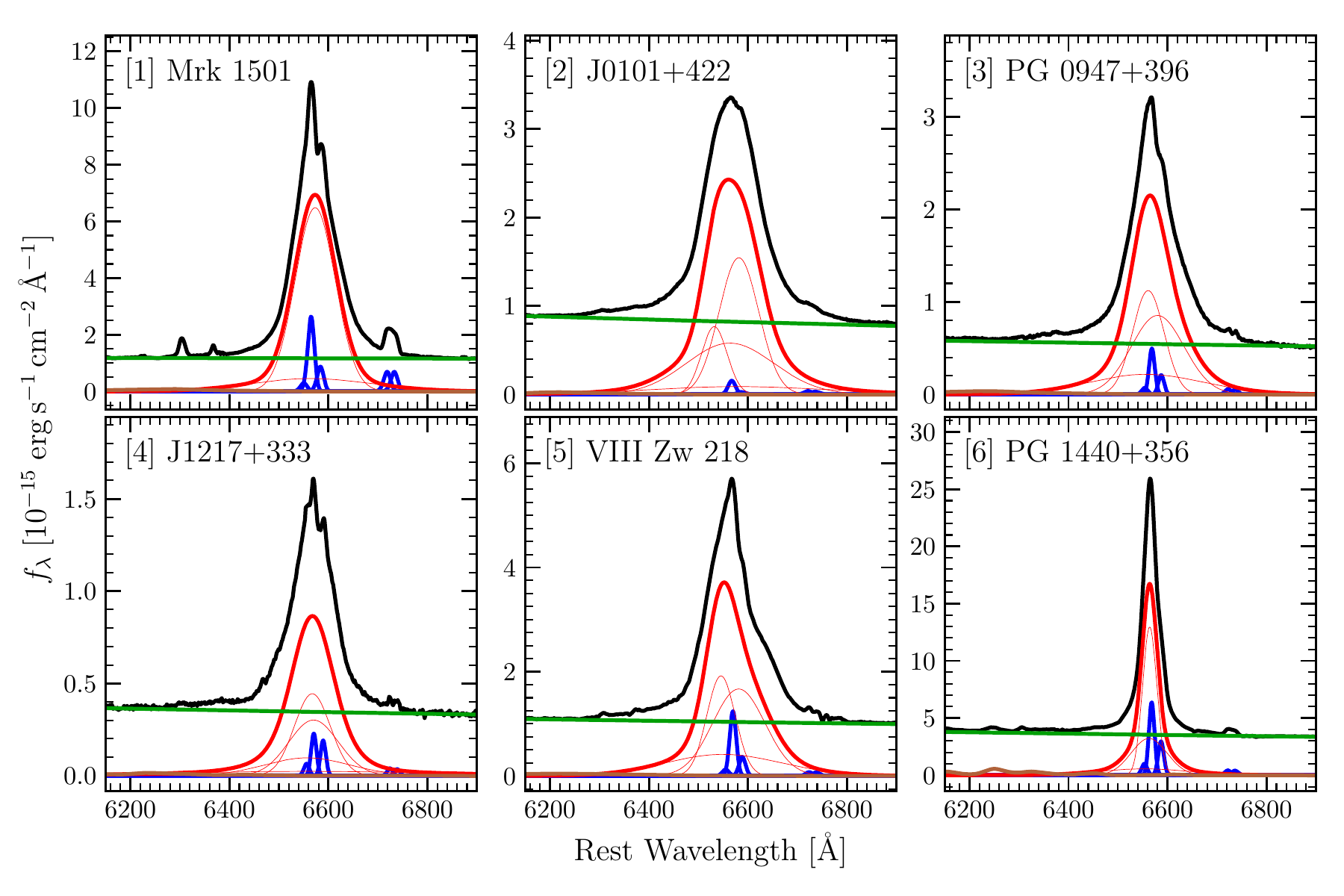}
\caption{Mean spectra of 6 objects around the \ha{} line and their best-fit models. Black represents the mean spectrum, green represents the power-law continuum, and brown represents the \ion{Fe}{2} model from \citet{Boroson&Green92}. Each blue line represents a narrow emission line. The thick red line shows the broad \ha{} model, whereas thin red lines show the individual Gaussian components of the model. \label{fig:decomp}}
\end{figure*}

To constrain the narrow line profile, we first modeled the \SII{}~$\lambda\lambda 6717,6731$ doublet. First, we fitted the wing of the broad \ha{} as a cubic polynomial in the windows of 6685-6708~\angstrom{} and 6760-6785~\angstrom{} . After subtracting the model of the broad \ha{} wing, each of the \SII{} lines was fitted with a single Gaussian profile using the maximum likelihood estimators. The acquired models of the \SII{} were then subtracted from the observed spectrum. Then, the narrow \ha{} and \NII{} lines were modeled as 1-2 Gaussian profiles, with shared shifts and widths among different lines, along with a sum of 2-4 independent Gaussian profiles for broad \ha{}. Upon fitting, we masked [\ion{O}{1}]$\lambda\lambda 6300,6364$ lines and, in the case of VIII~Zw~218, the telluric line residuals as well. We imposed a prior such that the narrow line shifts and widths follow normal distributions centered at the measurement from \SII{} with a standard deviation of $\sim 100\,\rm km\, s^{-1}$. Furthermore, we restricted the parameters to have the following bounds: 1. the velocity shift of any component is within $\pm 1600\,\mathrm{km\,s^{-1}}$, 2. the narrow line dispersion is smaller than $800\,\mathrm{km\,s^{-1}}$, and 3. the line dispersion of any Gaussian in broad line model is smaller than $30,000\,\mathrm{km\,s^{-1}}$, and is further restricted based on visual inspection of the spectrum. We adopted the maximum a posteriori estimator from MCMC samplings as our mean spectrum model. We also measured the FWHMs of the model line profiles and found their uncertainties based on Monte Carlo randomization for 1000 iterations. At each iteration, we added Gaussian random noise to the model parameters based on their measurement uncertainties and constructed a randomized profile. We measured the FWHMs of the randomized profiles and adopted their standard deviation as the uncertainty of the FWHM. The decomposed \ha{} and other narrow lines are shown in Figure~\ref{fig:decomp}, and the measurements after subtracting the instrumental resolving power, $R=650$, are summarized in Table~\ref{table:linewidth}. Note that the line widths listed here should only be used as a reference to the fit quality since the resolving power measured from the airglow lines can be underestimated, as noted in \S~\ref{ss:specred}.

\renewcommand{\arraystretch}{1.25}
\begin{deluxetable*}{rlcccccc}
\tablewidth{0.95\textwidth}
	\tablecolumns{8}
	\tablecaption{Line Width Measurements\label{table:linewidth}}
	\tablehead
	{
		&\colhead{Object}& \multicolumn{3}{c}{{Narrow}} &\multicolumn{3}{c}{{Broad \ha{}}}
		\\
		&\colhead{}&\colhead{$N$} &\colhead{$\sigma\, [\rm km\,s^{-1}]$} &\colhead{FWHM $[\rm km\,s^{-1}]$}&\colhead{$N$} &\colhead{$\sigma\, [\rm km\,s^{-1}]$} &\colhead{FWHM $[\rm km\,s^{-1}]$}
		\\
		&\colhead{(1)}    &\colhead{(2)}       &\colhead{(3)}&\colhead{(4)}&\colhead{(5)}&\colhead{(6)}&\colhead{(7)}
	}
	\startdata
1& Mrk~1501      & 2 & $287\pm159$ & $542\pm231$ & 2 & $2845\pm 465$ & $4532\pm 258$  \\
2& J0101$+$422   & 1 & $279\pm207$ & $657\pm489$ & 4 & $3862\pm1295$ & $5819\pm1624$  \\
3& PG~0947$+$396 & 1 & $176\pm207$ & $415\pm488$ & 3 & $3173\pm 808$ & $4553\pm1099$  \\
4& J1217$+$333   & 1 & $146\pm212$ & $344\pm500$ & 4 & $4828\pm4039$ & $4762\pm2040$  \\
5& VIII~Zw~218   & 1 & $226\pm187$ & $532\pm441$ & 3 & $3322\pm 709$ & $4788\pm 863$  \\
6& PG~1440$+$356 & 1 & $165\pm186$ & $388\pm438$ & 3 & $1896\pm 665$ & $1741\pm 288$  \\
	\enddata
    \tablecomments{Columns are (1) object name, (2) number of Gaussian components in the narrow line, (3) narrow line width in $\sigma$, (4) line width in full width at half maximum (FWHM), (5) number of Gaussian components in the broad line, (6) broad line width in $\sigma$, and (7) broad line width in full width at half maximum (FWHM). All values presented here are after correcting for the instrumental resolution, FWHM$_{\rm inst}=461\,\rm km\,s^{-1}$ or $\sigma_{\rm inst}=196\,\rm km\,s^{-1}$. The uncertainties shown here denote the standard deviation, where the uncertainty of $\sigma$ was derived using the analytical model, and the uncertainty of FWHM was derived with 1000 iterations of Monte Carlo randomization (see \S~\ref{s:specanal}).}
\end{deluxetable*}

To measure the broad \ha{} flux from each epoch, we modeled the spectra from individual nights using the mean spectrum model. We first modeled and subtracted the continuum and \ion{Fe}{2} lines using the same wavelength windows. We masked the [\ion{O}{1}] lines and telluric residuals as we did upon fitting the mean spectra. Then, assuming the narrow lines did not change over the period of observations, we subtracted the narrow line model obtained from the mean spectrum. After subtracting the continuum, \ion{Fe}{2}, and narrow lines, the residual spectrum contained the broad \ha{} line only. We constructed the prior for the multiple Gaussian models for broad \ha{} as follows. For the parameters that determined the shape of the \ha{} line (i.e., the flux ratio and 1st/2nd-moment differences between any pair of Gaussian components), we imposed Gaussian priors with the mean and the standard deviation from the mean spectrum model. We did not favor any specific value for the total flux of the \ha{} line and adopted a flat prior. For each spectrum, we calculated the maximum a posteriori estimators from MCMC samplings. Finally, the flux of the best-fit model was taken as the \ha{} flux of each epoch. Note that the uncertainty of the narrow emission line fitting does not affect the \ha{} lag measurements as we subtracted a constant flux of narrow emission lines in each epoch, which is expected to be non-varying during the campaign.


\section{Time Lag Measurements}\label{s:timelag}

\begin{figure*}[!thpb]
\centering
\includegraphics[width=0.95\textwidth]{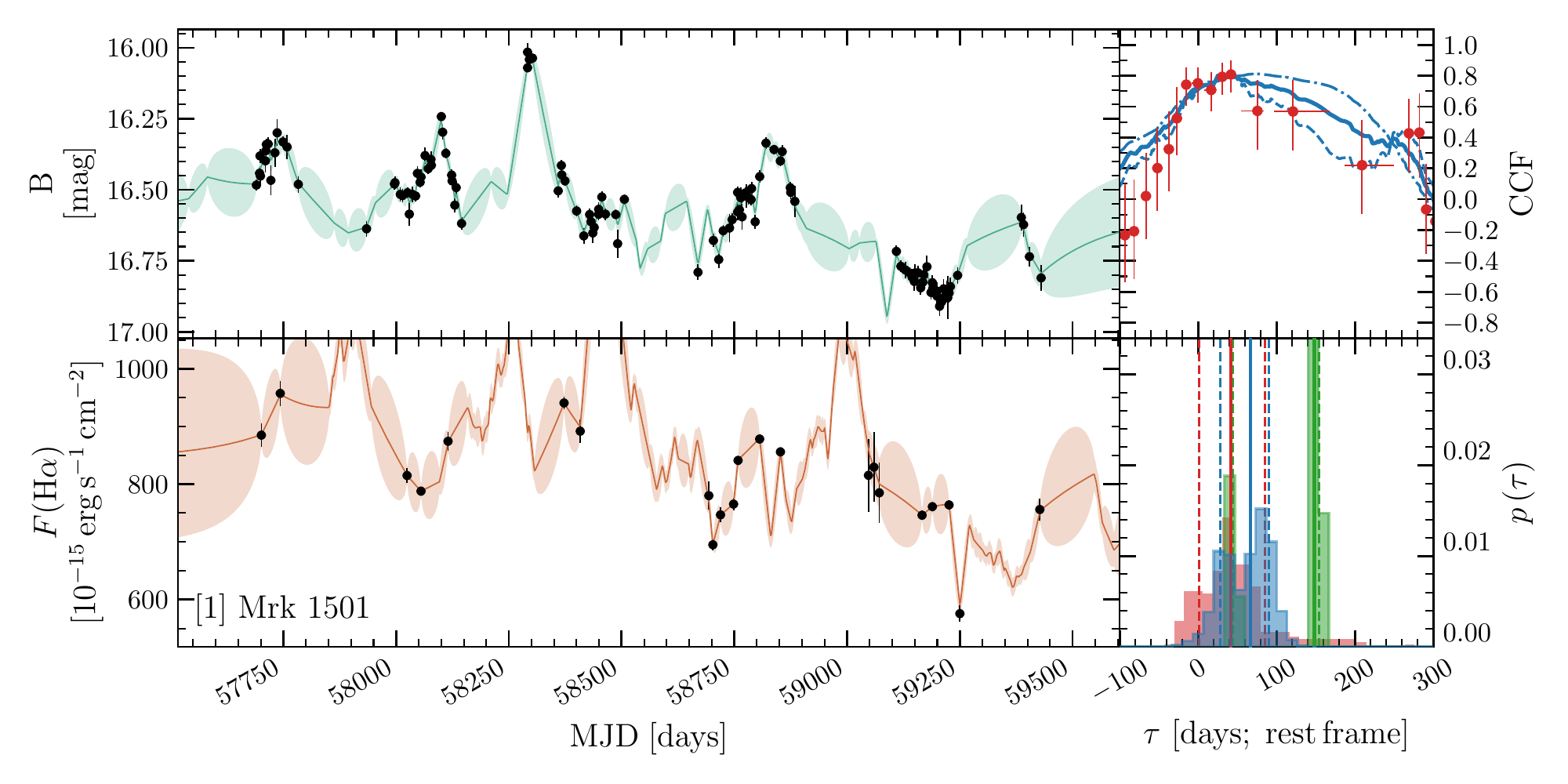}
\caption{Light curves and the time lag measurements of Mrk~1501.
\emph{Upper left}: B-band light curve, with the \jav{} model and its uncertainty shown as a solid line and shaded region.
\emph{Lower left}: Broad \ha{} light curve, with the \jav{} model shown similarly.
\emph{Upper right}: Blue lines indicate the ICCF, where the dashed line shows the ICCF upon interpolating the continuum, and the dash-dotted line shows the ICCF upon interpolating the line flux. The solid line represents the $z$-transformed average between two interpolations, as described in \S\ref{s:timelag}. Red dots indicate the \zdcf{}, where the horizontal error bar indicates the bin size of the \zdcf{}, and the vertical error bars indicate the standard deviation within the bin.
\emph{Lower right}: The time lag measurements. ICCF, \zdcf{}, and \jav{} are indicated by blue, red, and green colors, respectively. Vertical solid lines represent the median of each measurement, while dashed lines mark the 16th and 84th percentiles as 1-$\sigma$ uncertainties. \label{fig:obje}}
\end{figure*}
\begin{figure*}[!thpb]
\centering
\includegraphics[width=0.95\textwidth]{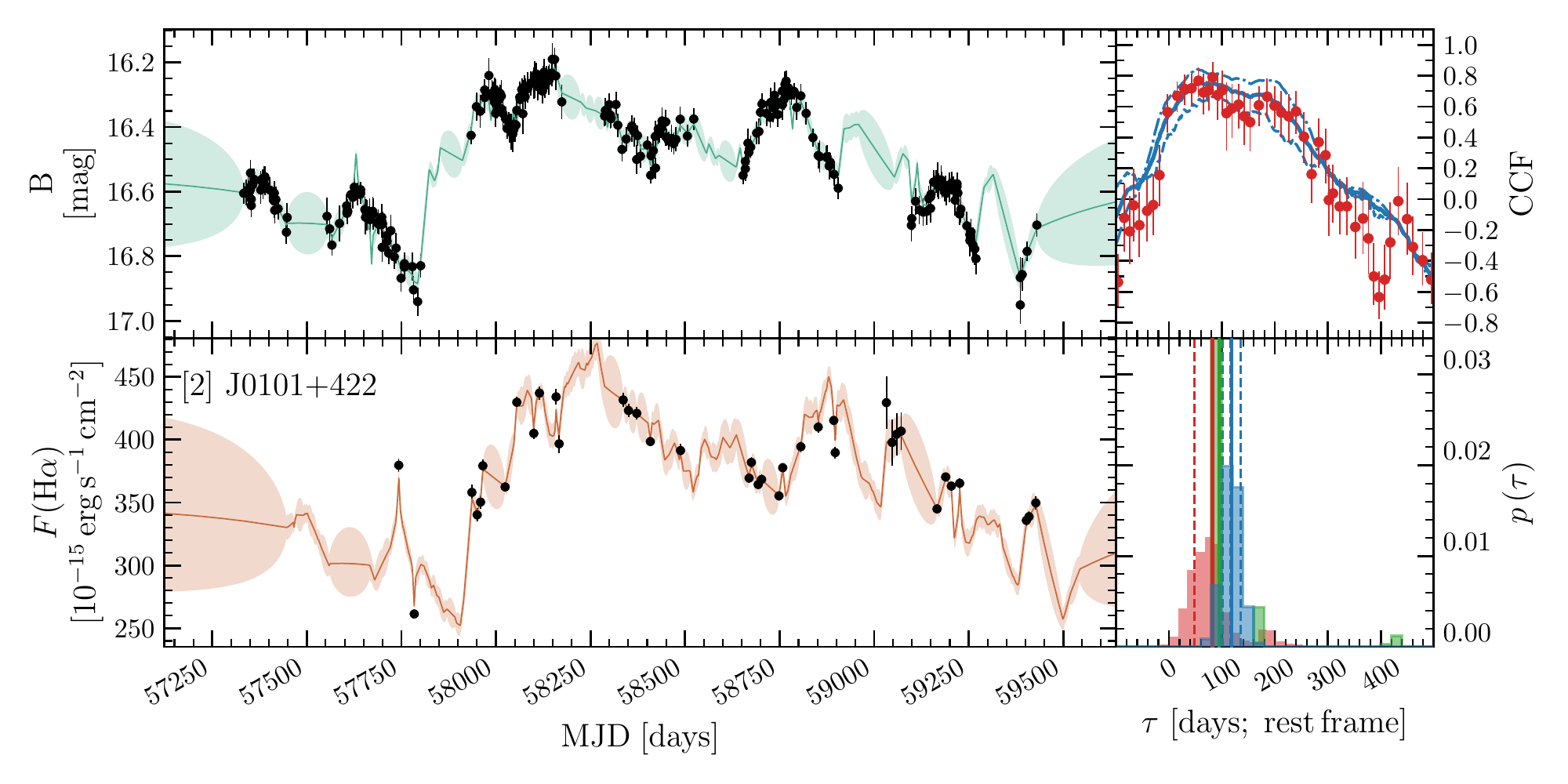}
\caption{Light curves and time lag measurements of J0101$+$422. The panels are the same as in Figure~\ref{fig:obje}.\label{fig:obja}}
\end{figure*}
\begin{figure*}[!thpb]
\centering
\includegraphics[width=0.95\textwidth]{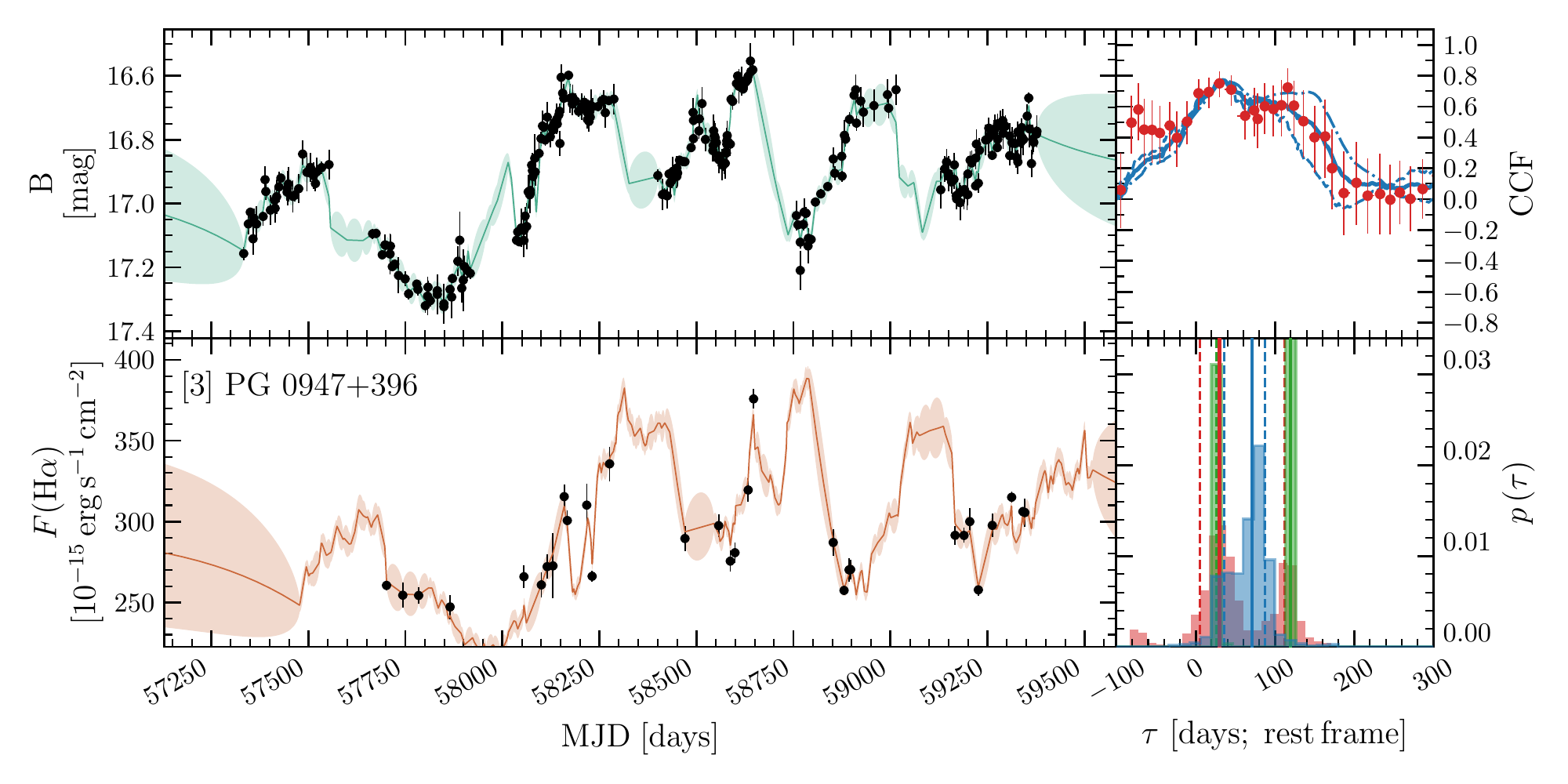}
\caption{Light curves and time lag measurements of PG~0947$+$396. The panels are the same as in Figure~\ref{fig:obje}.\label{fig:objb}}
\end{figure*}
\begin{figure*}[!thpb]
\centering
\includegraphics[width=0.95\textwidth]{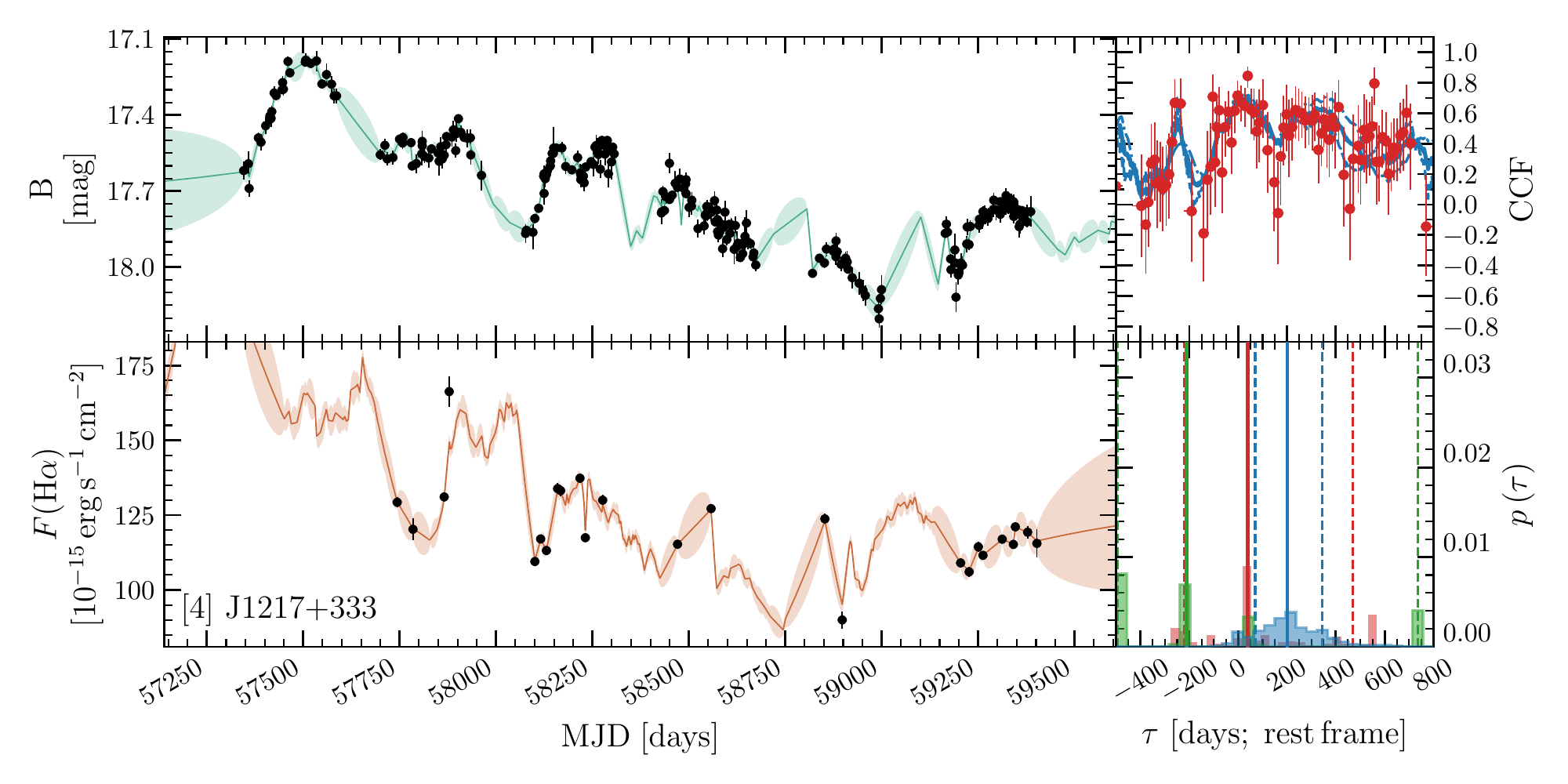}
\caption{Light curves and time lag measurements of J1217$+$333. The panels are the same as in Figure~\ref{fig:obje}.\label{fig:objf}}
\end{figure*}
\begin{figure*}[!thpb]
\centering
\includegraphics[width=0.95\textwidth]{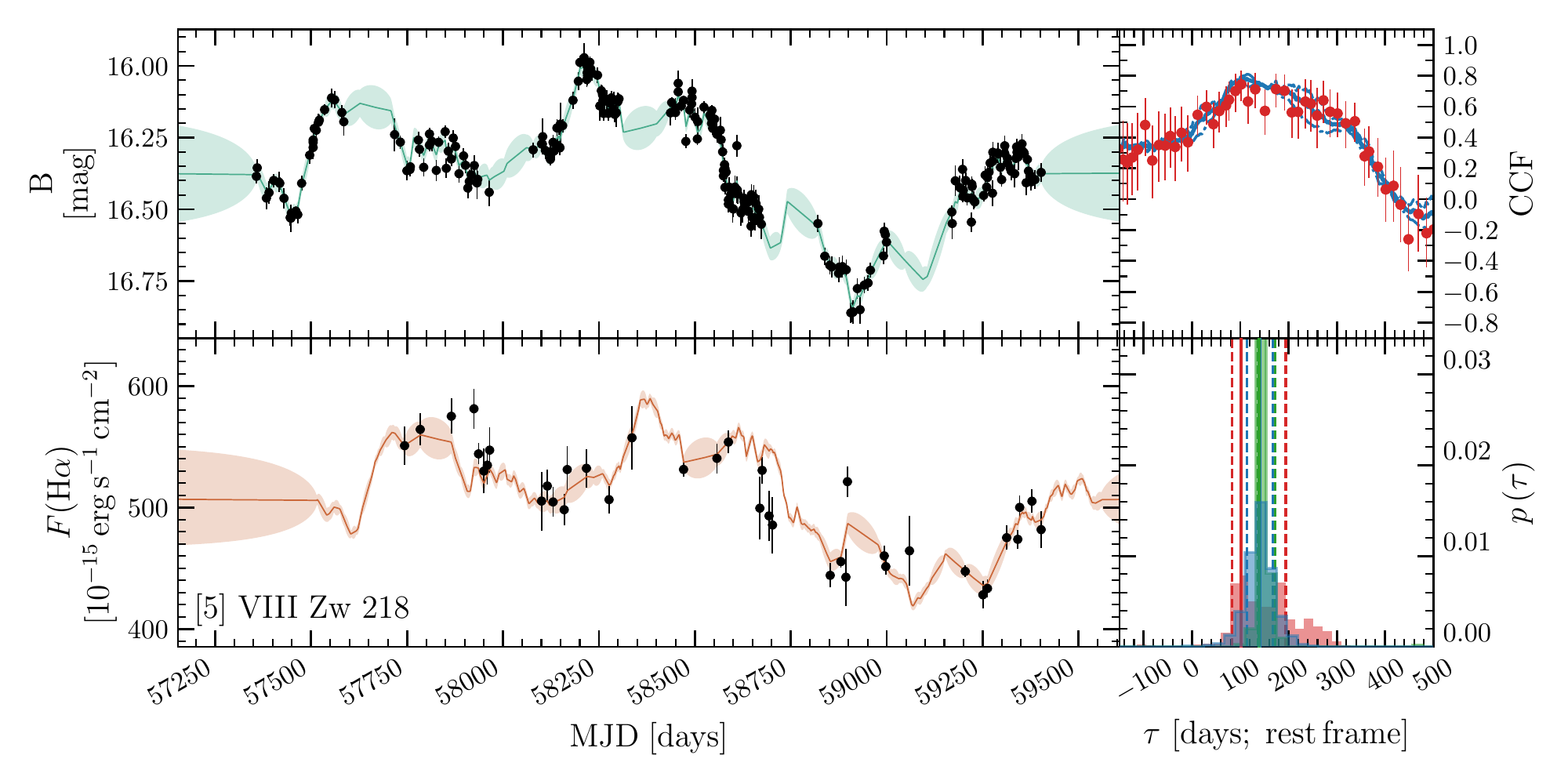}
\caption{Light curves and time lag measurements of VIII~Zw~218. The panels are the same as in Figure~\ref{fig:obje}.\label{fig:objc}}
\end{figure*}
\begin{figure*}[!thpb]
\centering
\includegraphics[width=0.95\textwidth]{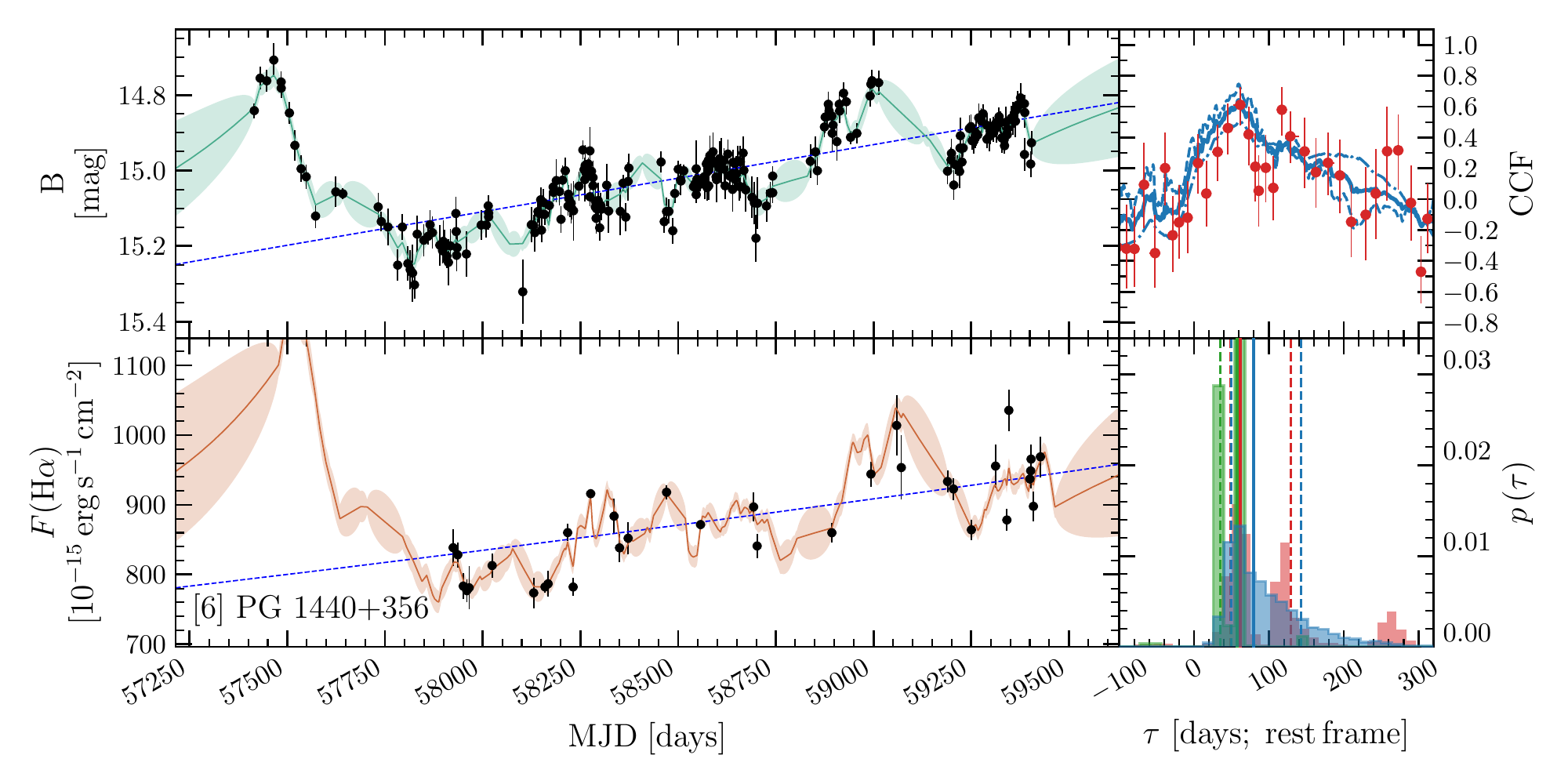}
\caption{Light curves and time lag measurements of PG~1440$+$356. The panels are the same as in Figure~\ref{fig:obje}, except in left panels, where trend fits for detrending are denoted as blue dashed lines.\label{fig:objd}}
\end{figure*}

We measured the time lags between the continuum and \ha{} line light curves using the interpolated cross-correlation function \citep[ICCF;][]{White&Peterson94} with a modified averaging scheme. We first converted the \ha{} line fluxes into magnitudes. After we calculated the continuum-interpolated ICCF and the line-interpolated ICCF, we took the Fisher transformation \citep{Fisher1921} on both one-sided ICCFs, averaged them, and then took the inverse transformation to obtain the $z$-transformed average ICCF. The ICCF centroid was calculated over the largest continuous interval containing the peak of the ICCF, where the ICCF values within the interval were larger than 80\% of the peak value. We performed 10,000 realizations of the flux randomization/random-subset selection \citep[FR/RSS; ][]{Peterson+98a, Peterson+04}. For each realization, we resampled each light curve and added random Gaussian noise to each flux value according to their measurement uncertainty. Duplicate points were averaged, and their uncertainties were divided by $\sqrt{n}$ to compensate for the duplication. The median of the distribution of the centroid and the central 68\% confidence interval was taken as the time lag measurement and the associated uncertainties. To check the consistency with the ICCF, we also calculated the time lag using other commonly-used methods, the $z$-transformed discrete correlation function \citep[\zdcf{}; ][]{Alexander97} and the \jav{} model \citep{Zu+11} Finally, we rated the quality of our lag measurements based on the lag differences among different methods as follows:
\begin{itemize}
\item Rating A if its ICCF lag, \zdcf{} lag, and \jav{} lag agree within 1-$\sigma$ and the maximum difference between them is within two months (60 days).
\item Rating B if its lag is measured (1-$\sigma$ above zero lag) with all 3 methods, while the maximum difference between them is larger than two months (60 days).
\item Rating C if its lag is not constrained or detected.
\end{itemize}
The measured lags are summarized in {Table~\ref{table:timelag}}. The light curves and the CCFs of individual objects are shown in Figures~\ref{fig:obje}, \ref{fig:obja}, \ref{fig:objb}, \ref{fig:objf}, \ref{fig:objc}, and \ref{fig:objd}.

\renewcommand{\arraystretch}{1.25}
\begin{deluxetable}{rlrrrc}[!th]
	\tablewidth{0.45\textwidth}
	\tablecolumns{6}
	\tablecaption{Rest-frame \ha{} Time Lag\label{table:timelag}}
	\tablehead
	{
		&\colhead{Object}& \colhead{ICCF} &\colhead{\zdcf} &\colhead{\jav} & \colhead{\thead{Quality\\Rating}}
		\\
		&\colhead{}&\colhead{[day]}&\colhead{[day]} &\colhead{[day]} &\colhead{}
		\\
		&\colhead{(1)}    &\colhead{(2)}       &\colhead{(3)}&\colhead{(4)} &\colhead{(5)}
	}
	\startdata
1&Mrk~1501      & \valerrud{ 67}{ 24}{ 38} & \valerrud{ 42}{ 44}{ 40} & \valerrud{ 148}{  6}{104} & B \\
2&J0101$+$422   & \valerrud{118}{ 17}{ 17} & \valerrud{ 82}{ 34}{ 35} & \valerrud{  95}{  1}{ 11} & A \\
3&PG~0947$+$396 & \valerrud{ 71}{ 16}{ 35} & \valerrud{ 30}{ 83}{ 25} & \valerrud{ 120}{  1}{ 93} & B \\
4&J1217$+$333   & \valerrud{201}{143}{132} & \valerrud{ 39}{432}{258} & \valerrud{-211}{946}{284} & C \\
5&VIII~Zw~218   & \valerrud{140}{ 26}{ 26} & \valerrud{102}{ 92}{ 19} & \valerrud{ 139}{ 33}{  3} & A \\
6&PG~1440$+$356 & \valerrud{ 80}{ 63}{ 30} & \valerrud{ 62}{ 68}{ 13} & \valerrud{  61}{  3}{ 23} & A \\
	\enddata
    \tablecomments{Columns are (1) object identifier, (2) ICCF/CCCD lag, (3) \zdcf{} lag, (4) \jav{} lag, and (5) quality rating as described in \S~\ref{s:timelag}. Uncertainties shown here are the 68\% central confidence intervals taken from the posterior distribution.}
\end{deluxetable}

Here, we describe individual measurements of six targets. For Mrk~1501, we measured \valerrud{ 68}{ 23}{ 41}~days of time lag. However, the CCFs at time lags between 100 days and 200 days were relatively unexplored due to the large seasonal gaps. This is reflected in the large difference of ICCFs using different interpolation methods, as well as the lack of points in the \zdcf{}. The primary peak of the \jav{} lag distribution falls in this seasonal gap, which questions the accuracy of the \jav{} lag for this specific case. We assess the lag of this object as rating B.

For J0101$+$422, we measured \valerrud{117}{17}{18}~days of the time lag. This is consistent with its \zdcf{} and \jav{} lags. We assess the lag of this object as rating A.

For PG~0947$+$396, we measured \valerrud{71}{16}{34}~days of the time lag. While this is consistent with the \zdcf{} and \jav{} lags, they both showed bimodality in their lag, where \zdcf{} preferred the smaller mode and \jav{} preferred the larger mode. On the other hand, the ICCF lag captured the average between the two lags. We assess the lag of this object as rating B.

For J1217$+$333, we assess that our lag measurements are not reliable. \jav{} is unconstrained, and the \zdcf{} lag is consistent with 0 within its 68\% confidence interval. While we measured \valerrud{199}{137}{132}~days from the ICCF, this is likely to be the average of the window size we used to find the lag. Supporting this, the ICCF and \zdcf{} both show multiple modes in the given window. Moreover, it is far larger than what is measured using \hb{} \citep{samp_hb_inprep}. We assess the lag of this object as rating C and exclude it from further analysis.

VIII~Zw~218 showed a single, clean peak in CCFs, and we measured \valerrud{140}{25}{25}~days of the time lag. This value is consistent with \zdcf{} and \jav{} lags. We assess the lag of this object as rating A.

We had to detrend the light curves of PG~1440$+$356 before measuring the time lag since the \ha{} light curve showed a monotonic increase in the \ha{} light curve. Without detrending, the CCF values at any lag were higher than 0.4, rendering the lag and its uncertainty measurements unreliable. After detrending, we measured \valerrud{79}{68}{29}~days of time lag. This value is consistent with \zdcf{} and \jav{} lags. We note that both ICCF and \zdcf{} are skewed toward higher lag values, and this is reflected in the measurements. We assess the lag of this object as rating A.

We find that the uncertainties of our ICCF-based lag measurements range from 10\% to 70\%, a considerable portion of which stems from the number of epochs in the light curve. The lack of distinct variability features in the light curve also increases the uncertainty. However, we note that the relative uncertainties of our lag measurements are comparable to those reported in the literature. For example, 95\% of the lag measurements we collected in Table~\ref{table:fitdata} have uncertainty between 8\%-70\%.


\section{The Size-Luminosity Relation}\label{s:sl}
\subsection{Determining the \texorpdfstring{\ha{}}{} BLR size-luminosity relation}
To compare the BLR size of \ha{} with \ha{} luminosity, we compiled a sample of \ha{} lags and luminosities from this work and the literature \citep{Kaspi+00, Bentz+10, Barth+11a, Grier+17, Sergeev+17, Cho+20, Feng+21, Li+22}. While we tried to include as many AGNs as possible, the following objects were excluded. First, \citet{Grier+17} provided quality ratings for the time lags, and 4 objects (out of 18 with \ha{} lags measured) with ratings of 1 or 2 were excluded. Additionally, we note that \ha{} time lags of Ark~564 \citep{Shapovalova+12} and NGC~7469 \citep{Shapovalova+17} were measured. However, the \ha{} lag of Ark~564 was consistent with 0 delay within the error bar, while the CCF between \ha{} and the continuum light curves of NGC~7469 exhibited multiple peaks with $r_{\rm max}$ values smaller than 0.5. Thus, we did not include these two objects in our analysis. We adopted the time lag values and their 1-$\sigma$ uncertainties from each paper. For objects presented in this paper, we adopted the ICCF time lag values and their 1-$\sigma$ errors. The lags presented in the observed frame were divided by $1+z$ to convert them into the rest-frame lags. We obtained the broad \ha{} luminosities from the fluxes by multiplying $4\pi {d_L}^2$, where $d_L$ is the luminosity distance. We ignored the systematic differences in methods of measuring flux between the papers, which are discussed in \S~\ref{ss:difflum}. We also collected the AGN continuum luminosities at 5100\angstrom{} for the sample. We applied galactic extinction correction based on the galactic extinction map by \citet{SandF11} and the extinction curve by \citet{CCM89}. The corrected luminosities, as well as their time lags, are listed in Table~\ref{table:fitdata}.
\renewcommand{\arraystretch}{1.25}
\startlongtable
\begin{deluxetable*}{lccrrc}
\tablewidth{0.99\textwidth}
\tablecolumns{6}
\tablecaption{The Time Lags and Luminosities \label{table:fitdata}}
\tablehead
{
\colhead{Object}& \colhead{$\log_{10} L_{\rm H\alpha}$} & \colhead{$\log_{10} \lambda L_\lambda \left({\rm 5100\angstrom{}}\right)$} &\colhead{$\tau_{\rm H\alpha}$} &\colhead{$\tau_{\rm H\beta}$} & \colhead{References}
\\
\colhead{}&\colhead{[$\rm erg\,s^{-1}$]}&\colhead{[$\rm erg\,s^{-1}$]}&\colhead{[day]} &\colhead{[day]} & \colhead{}
\\
\colhead{(1)}    &\colhead{(2)}       &\colhead{(3)}&\colhead{(4)}&\colhead{(5)}&\colhead{(6)}
}
\startdata
Mrk~1501 & $43.15\pm0.02$ & $44.14\pm0.02$ & \valerrud{67}{24}{38} & \valerrud{12}{8}{9} & 1 \\
PG 0026+129 & $43.53\pm0.04$ & $44.98\pm0.06$ & \valerrud{116}{25}{27} & \valerrud{109}{25}{32} & 2 \\
PG 0052+251 & $43.70\pm0.05$ & $44.92\pm0.08$ & \valerrud{183}{57}{38} & \valerrud{86}{26}{27} & 2 \\
J0101$+$422 & $43.58\pm0.01$ & $44.89\pm0.01$ & \valerrud{118}{17}{17} & \valerrud{76}{13}{12} & 1 \\
PG 0804+761 & $43.72\pm0.02$ & $44.90\pm0.08$ & \valerrud{175}{18}{15} & \valerrud{137}{24}{22} & 2 \\
NGC 2617 & $41.42\pm0.01$ & - & \valerrud{6.9}{1.6}{0.8} & \valerrud{5.4}{1.0}{1.1} & 6 \\
PG 0844+349 & $42.98\pm0.03$ & $44.31\pm0.04$ & \valerrud{37}{15}{15} & \valerrud{12}{13}{10} & 2 \\
PG~0947$+$396 & $43.52\pm0.02$ & $44.71\pm0.01$ & \valerrud{71}{16}{35} & \valerrud{37}{10}{11} & 1 \\
Mrk 142 & $42.08\pm0.03$ & $43.57\pm0.04$ & \valerrud{2.8}{1.2}{0.9} & \valerrud{2.7}{0.7}{0.8} & 3 \\
SBS 1116+583A & $41.25\pm0.03$ & $42.46\pm0.03$ & \valerrud{4.0}{1.4}{1.0} & \valerrud{2.3}{0.6}{0.5} & 3 \\
Arp 151 & $41.52\pm0.05$ & $42.44\pm0.05$ & \valerrud{7.8}{1.0}{1.0} & \valerrud{4.0}{0.5}{0.7} & 3 \\
Mrk 1310 & $41.34\pm0.03$ & $42.60\pm0.03$ & \valerrud{4.5}{0.7}{0.6} & \valerrud{3.7}{0.6}{0.6} & 3 \\
NGC 4151 & $41.70\pm0.02$ & $42.66\pm0.09$ & \valerrud{7.6}{1.9}{2.6} & \valerrud{6.2}{1.4}{1.1} & 7 \\
PG 1211+143 & $43.54\pm0.05$ & $44.77\pm0.07$ & \valerrud{107}{35}{42} & \valerrud{95}{29}{41} & 2 \\
Mrk 202 & $41.13\pm0.03$ & $42.70\pm0.02$ & \valerrud{22}{1}{4} & \valerrud{3.1}{1.7}{1.1} & 3 \\
NGC 4253 & $41.62\pm0.02$ & $42.80\pm0.02$ & \valerrud{25}{1}{1} & \valerrud{6.2}{1.6}{1.2} & 3 \\
NGC 4395 & $38.45\pm0.00$ & $39.76\pm0.01$ & \valerrud{0.058}{0.010}{0.010} & - & 5 \\
PG 1226+023 & $44.60\pm0.03$ & $45.93\pm0.05$ & \valerrud{444}{56}{55} & \valerrud{330}{101}{83} & 2 \\
PG 1229+204 & $42.87\pm0.04$ & $44.07\pm0.05$ & \valerrud{67}{37}{43} & \valerrud{34}{30}{17} & 2 \\
NGC 4748 & $41.68\pm0.03$ & $42.79\pm0.02$ & \valerrud{7.5}{3.0}{4.6} & \valerrud{5.5}{1.6}{2.2} & 3 \\
VIII~Zw~218 & $43.29\pm0.02$ & $44.53\pm0.01$ & \valerrud{140}{26}{26} & \valerrud{63}{16}{15} & 1 \\
PG 1307+085 & $43.68\pm0.04$ & $44.84\pm0.04$ & \valerrud{155}{81}{126} & \valerrud{94}{40}{100} & 2 \\
PG 1351+640 & $43.16\pm0.02$ & $44.73\pm0.04$ & \valerrud{227}{149}{72} & - & 2 \\
SDSS J140812.09+535303.3 & $41.68\pm0.04$ & $43.16\pm0.00$ & \valerrud{7.2}{4.8}{5.6} & \valerrud{9.0}{5.6}{3.8} & 4 \\
SDSS J140915.70+532721.8 & $42.27\pm0.03$ & $43.40\pm0.00$ & \valerrud{33}{14}{10} & - & 4 \\
SDSS J141018.04+532937.5 & $42.17\pm0.04$ & $43.56\pm0.01$ & \valerrud{23}{13}{8} & \valerrud{14}{4}{6} & 4 \\
SDSS J141041.25+531849.0 & $42.49\pm0.02$ & $43.79\pm0.01$ & \valerrud{12}{8}{7} & \valerrud{11}{7}{7} & 4 \\
SDSS J141123.42+521331.7 & $42.62\pm0.03$ & $44.12\pm0.01$ & \valerrud{13}{10}{14} & \valerrud{6.5}{8.8}{5.4} & 4 \\
PG 1411+442 & $43.40\pm0.02$ & $44.59\pm0.04$ & \valerrud{95}{37}{34} & \valerrud{108}{66}{65} & 2 \\
SDSS J141151.78+525344.1 & $42.68\pm0.06$ & $44.15\pm0.01$ & \valerrud{55}{4}{5} & - & 4 \\
SDSS J141324.28+530527.0 & $42.44\pm0.05$ & $43.91\pm0.00$ & \valerrud{45}{14}{11} & \valerrud{22}{11}{11} & 4 \\
SDSS J141625.71+535438.5 & $42.71\pm0.01$ & $43.95\pm0.00$ & \valerrud{33}{19}{17} & \valerrud{17}{6}{7} & 4 \\
SDSS J141645.15+542540.8 & $41.85\pm0.07$ & $43.24\pm0.01$ & \valerrud{9.6}{4.5}{3.0} & \valerrud{6.5}{2.7}{1.8} & 4 \\
SDSS J141645.58+534446.8 & $42.03\pm0.05$ & $43.64\pm0.01$ & \valerrud{18}{7}{8} & \valerrud{9.7}{4.0}{4.0} & 4 \\
SDSS J141751.14+522311.1 & $41.97\pm0.03$ & $42.80\pm0.01$ & \valerrud{11}{6}{5} & - & 4 \\
NGC 5548 & $42.06\pm0.03$ & $43.10\pm0.03$ & \valerrud{11}{1}{1} & \valerrud{4.2}{0.9}{1.3} & 3 \\
SDSS J142038.52+532416.5 & $42.08\pm0.03$ & $43.46\pm0.00$ & \valerrud{20}{15}{15} & \valerrud{27}{8}{14} & 4 \\
SDSS J142039.80+520359.7 & $42.57\pm0.03$ & $44.10\pm0.01$ & \valerrud{18}{6}{16} & \valerrud{5.1}{6.4}{8.5} & 4 \\
SDSS J142135.90+523138.9 & $41.96\pm0.06$ & $43.44\pm0.00$ & \valerrud{7.2}{3.4}{5.6} & \valerrud{1.0}{3.8}{4.2} & 4 \\
PG 1426+015 & $43.40\pm0.03$ & $44.68\pm0.07$ & \valerrud{83}{42}{48} & \valerrud{106}{45}{63} & 2 \\
PG~1440$+$356 & $43.08\pm0.03$ & $44.63\pm0.00$ & \valerrud{80}{63}{30} & \valerrud{51}{17}{21} & 1 \\
PG 1613+658 & $43.63\pm0.03$ & $44.95\pm0.05$ & \valerrud{38}{35}{19} & \valerrud{39}{18}{20} & 2 \\
PG 1617+175 & $43.24\pm0.03$ & $44.46\pm0.08$ & \valerrud{100}{28}{33} & \valerrud{70}{27}{37} & 2 \\
3C 390.3 & $42.95\pm0.02$ & $44.02\pm0.01$ & \valerrud{153}{14}{14} & \valerrud{84}{8}{8} & 8 \\
Zw 229-015 & $41.47\pm0.00$ & $42.65\pm0.05$ & \valerrud{5.1}{0.8}{1.1} & \valerrud{3.9}{0.7}{0.9} & 9 \\
NGC 6814 & $41.02\pm0.03$ & $42.10\pm0.03$ & \valerrud{9.5}{1.9}{1.6} & \valerrud{6.6}{0.9}{0.9} & 3 \\
PG 2130+099 & $43.19\pm0.03$ & $44.39\pm0.04$ & \valerrud{223}{50}{26} & \valerrud{177}{128}{25} & 2 \\
\enddata
\tablecomments{Columns are (1) object identifier, (2) luminosity of the broad \ha{} line, (2) continuum luminosity at 5100 $\rm\angstrom{}$, (4) broad \ha{} lag, (5) broad \hb{} lag, and (6) the time lag reference. All luminosity values are corrected for the galactic extinction based on the extinction value by \citet{SandF11} and the extinction curve by \citet{CCM89}. Time lag values are presented in the rest frame. Uncertainties shown here are the 68\% confidence intervals.}
\tablerefs{
1.~This Work with the continuum luminosity and \hb{} time lag by \citet{samp_hb_inprep},
2.~\citet{Kaspi+00},
3.~\citet{Bentz+10} with continuum luminosity by \citet{Bentz+09} and the host correction by \citet{Park+12},
4.~\citet{Grier+17} with continuum luminosity by \citet{Shen+15},
5.~\citet{Woo+19a} and \citet{Cho+20}, with the broad \ha{} luminosity by \citet{Cho+21},
6.~\citet{Feng+21},
7.~\citet{Li+22},
8.~\citet{Sergeev+17},
9.~\citet{Barth+11a} with continuum luminosity by \citet{Barth+15}.
}
\end{deluxetable*}

\begin{figure*}[!t]
\centering
\includegraphics[width=0.90\textwidth]{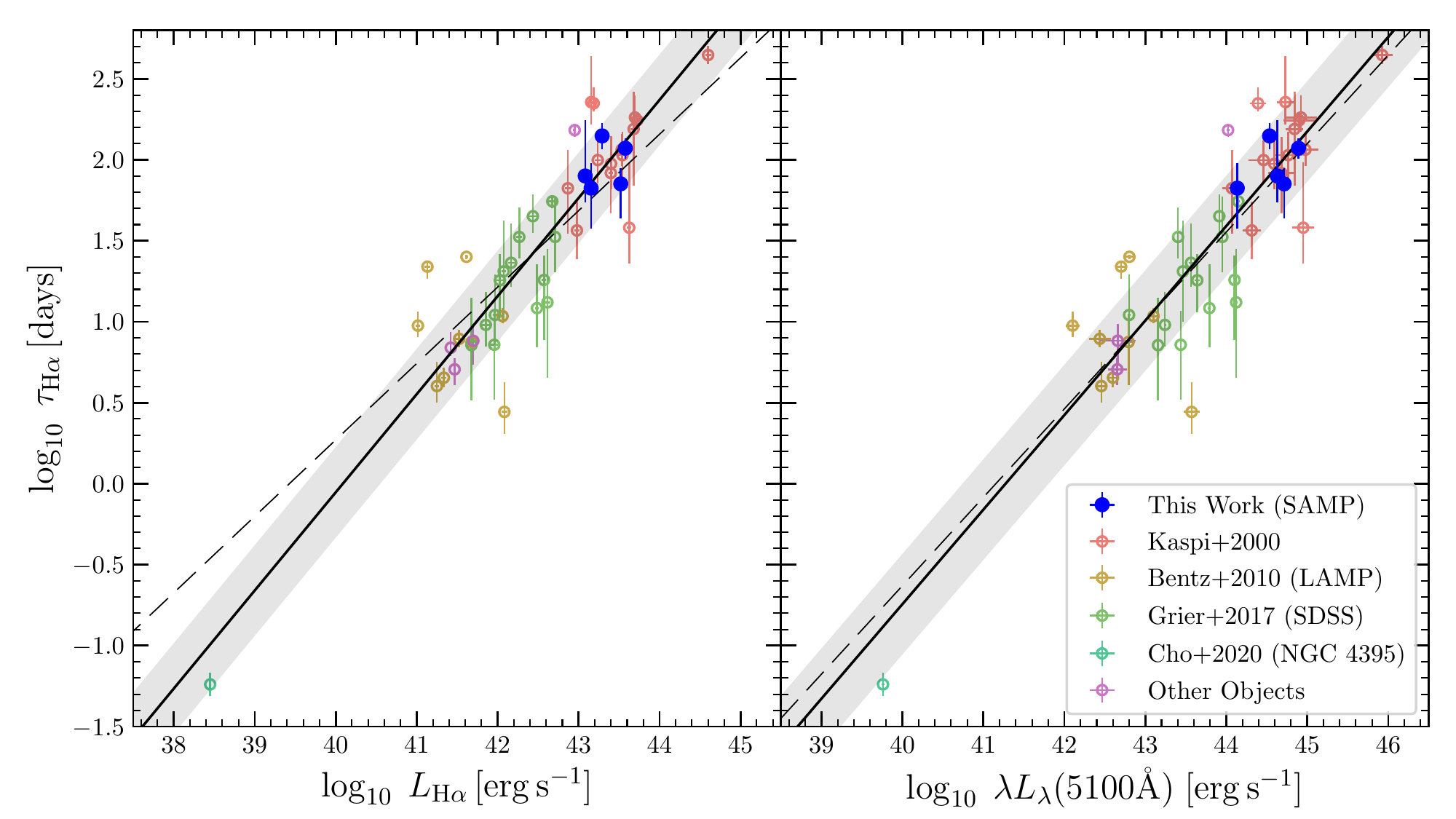}
\caption{\emph{Left}: The \ha{} time lag against the \ha{} line luminosity. The best-fit relation is denoted with a thick solid line, with the shaded region representing the 1-$\sigma$ intrinsic scatter. The dashed line represents the estimated \hb{} lag from the \ha{} luminosity by combining Equation 1 from \citet{Greene&Ho05} and Equation 2 from \citet{Bentz+13} with \emph{Clean+ExtCorr} fit parameters (See \S\ref{ss:mass}). \emph{Right}: The \ha{} time lag against the continuum luminosity at 5100\angstrom{}. The best-fit relation is denoted with a thick solid line, with the shaded area representing the 1-$\sigma$ intrinsic scatter. The dashed line represents the estimated \hb{} lag from the continuum luminosity using Equation 2 from \citet{Bentz+13} with \emph{Clean+ExtCorr} fit parameters. \label{fig:slrel}}
\end{figure*}

We model a generic relation between two variables $x$ and $y$ as
\begin{equation}
\log_{10} y = K + \beta \log_{10} x \pm \sigma_{\rm int}\label{eq:genericlin}
\end{equation}
where $\sigma_{\rm int}$ denotes the intrinsic scatter of the relation. To construct the size-luminosity relation of the \ha{} BLR, we chose $y$ to be $\tau_{\rm H\alpha}/30\,[{\rm days}]$, with $x$ being either $L_{\rm H\alpha}/10^{42}\,[{\rm erg\,s^{-1}}]$ or $\lambda L_{\lambda} \left({\rm 5100\angstrom{}}\right)/10^{44}\,[{\rm erg\,s^{-1}}]$. Here, variables are normalized to values close to the median of the sample to minimize the posterior correlation of $K$ and $\beta$. We fit this relation using the \texttt{Python} implementation of the \texttt{LINMIX\_ERR} algorithm \citep{Kelly07}\footnote{\url{https://github.com/jmeyers314/linmix}}. Since this code does not handle different values for upper and lower uncertainties, we took the mean of two uncertainties if both were present. The best-fit parameters are presented in Table~\ref{table:slfitpar}.

We should note that NGC~4395 has an extremely low luminosity, while PG~1226+023 has an extremely high luminosity compared to other AGNs in our sample. We tested whether removing these outliers from the sample would alter the fit. As presented in Table~\ref{table:slfitpar}, we found that excluding or including NGC~4395 and PG~1226+023 yields similar best fits, with the largest difference between parameters within the 1-$\sigma$ boundary. Thus, we present the best fit for the size-luminosity relation of the \ha{} BLR to be
\begin{equation}
\begin{aligned}
\log_{10}&\left(\frac{\tau_{\rm H\alpha}}{1[{\rm day}]}\right) = \left(1.16\pm 0.05\right)
\\& + \left(0.61\pm 0.04\right)\log_{10}\left(\frac{L_{\rm H\alpha}}{10^{42}\,[{\rm erg\,s^{-1}}]} \right)\label{eq:slfit}
\end{aligned}
\end{equation}
with $\sigma_{\rm int} = 0.28\pm 0.03$, and
\begin{equation}
\begin{aligned}
\log_{10}&\left(\frac{\tau_{\rm H\alpha}}{1[{\rm day}]}\right) = \left(1.59\pm 0.05\right)
\\& + \left(0.58\pm 0.04\right)\log_{10}\left(\frac{\lambda L_{\lambda} \left({\rm 5100\angstrom{}}\right)}{10^{44}\,[{\rm erg\,s^{-1}}]}\right)\label{eq:slfit_cont}
\end{aligned}
\end{equation}
where $\sigma_{\rm int} = 0.31\pm 0.03$. The best fit, as well as the objects used to derive the fit, are plotted in Figure~\ref{fig:slrel}.

\renewcommand{\arraystretch}{1.25}
\begin{deluxetable*}{ccccrrr}
\tablewidth{0.95\textwidth}
\tablecolumns{7}
\tablecaption{The Best-fit Parameters\label{table:slfitpar}}
\tablehead
{
    \colhead{$x$}& \colhead{$y$} &\colhead{\makecell{NGC 4395\\PG 1226+023}} & \colhead{$N$} & \colhead{$K$} & \colhead{$\beta$} & \colhead{$\sigma_{\rm int}$}
    \\
    \colhead{(1)} &\colhead{(2)} &\colhead{(3)}&\colhead{(4)}  &\colhead{(5)}&\colhead{(6)}&\colhead{(7)}
}
\startdata
$L_{\rm H \alpha}/10^{42}\,[{\rm erg\,s^{-1}}]$ & $\tau_{\rm H \alpha}/30\,[{\rm days}]$ & Included & $47$ & $-0.32\pm0.05$ & $0.61\pm0.04$ & $0.28\pm0.03$\\
& & Excluded & $45$ & $-0.32\pm0.05$ & $0.58\pm0.05$ & $0.28\pm0.03$\\
$\lambda L_{\lambda} \left({\rm 5100\angstrom{}}\right)/10^{44}\,[{\rm erg\,s^{-1}}]$ & $\tau_{\rm H \alpha}/30\,[{\rm days}]$ & Included & $46$ & $0.11\pm0.05$ & $0.58\pm0.04$ & $0.31\pm0.03$\\
& & Excluded & $44$ & $0.12\pm0.05$ & $0.54\pm0.06$ & $0.31\pm0.04$\\
$\lambda L_{\lambda} \left({\rm 5100\angstrom{}}\right)/10^{44}\,[{\rm erg\,s^{-1}}]$ & $L_{\rm H \alpha}/10^{42}\,[{\rm erg\,s^{-1}}]$ & Included & $46$ & $0.71\pm0.03$ & $0.97\pm0.03$ & $0.18\pm0.02$\\
&  & Excluded & $44$ & $0.71\pm0.03$ & $0.94\pm0.04$ & $0.19\pm0.02$\\
$\tau_{\rm H \alpha}/30\,[{\rm days}]$ & $\tau_{\rm H \beta}/30\,[{\rm days}]$ & PG 1226 only & $42$ & $-0.23\pm0.04$ & $1.01\pm0.07$ & $0.23\pm0.03$\\
&  & Excluded & $41$ & $-0.23\pm0.04$ & $1.00\pm0.07$ & $0.24\pm0.03$\\
\enddata
\tablecomments{Columns are (1) independent variable, (2) dependent variable, (3) inclusion of NGC 4395 and PG 1226+023 to the fit, (4) size of the subset, (5), (6), and (7) the best fit parameters to the model described by Eq.~\ref{eq:genericlin}. The best-fit parameters and their uncertainties are given in the median and the standard deviation.}
\end{deluxetable*}

\subsection{Systematic differences in luminosities}\label{ss:difflum}
We note that the narrow line fluxes were handled differently in the literature when measuring the \ha{} luminosity. For example, in their work, \citet{Cho+21}, \citet{Feng+21}, and \citet{Li+22} modeled the broad \ha{} line and narrow line components separately with multiple Gaussians. Similarly, \citet{Grier+17} modeled the broad line after the narrow lines were modeled and removed using high-pass filtering, which was described by \citet{Shen+16}. In contrast, \citet{Kaspi+00}, \citet{Bentz+10}, \citet{Barth+11a}, and \citet{Sergeev+17} integrated the flux over certain wavelength windows without removing the narrow lines of \ha{}, \NII{}, and even \SII{} depending on the object. In principle, direct integration leads to inaccurate measurements by including narrow line fluxes while excluding the wing of broad \ha{}. These two effects can work in the opposite way such that they could cancel each other out at a certain level. We assessed that the effect of the \ha{} wing seems negligible for the objects we included because all aforementioned references chose a sufficiently large interval for direct integration. Moreover, the narrow line fluxes seem negligible compared to their strong \ha{} luminosities for high luminosity objects that \citet{Kaspi+00} or \citet{Sergeev+17} presented. However, visually inspecting the spectra presented by \citet{Bentz+10} reveals that the \ha{} fluxes they measured do include a substantial portion of narrow line fluxes, resulting in a larger scatter.

The measurements of the continuum luminosity suffer from a similar issue as well. Specifically, \citet{Kaspi+00} and \citet{Sergeev+17} did not remove the host galaxy contamination when measuring the continuum flux at 5100\angstrom{}. However, the host contamination would be negligible for these high-luminosity AGNs. For instance, \citet{Jalan+23} proposed an empirical correction for host contamination based on the total luminosity at 5100\angstrom{}. According to their Equation 2, on average, 30\% of the continuum luminosities of 15 AGNs in the aforementioned references can be attributed to their host galaxy starlight. This corresponds to the bias of 0.15~dex, which is far smaller than the intrinsic scatter of our fit. However, we do not use their empirical correction because of its large scatter.

Despite the aforementioned issues, we do not observe any systematic deviation from the size-luminosity relations of the AGNs with inaccurate luminosity measurements. Thus, we conclude that neither the narrow lines nor the host contamination induced bias in determining the size-luminosity relation, albeit the intrinsic scatter could be overestimated.

For consistency check, we investigate whether the \ha{} BLR size–luminosity relation changes by excluding our new measurements. We find that the best-fit slope remains the same and the intrinsic scatter increases slightly (i.e., $0.29\pm0.03$ dex), presumably due to the small size of our new sample. However, the RMS scatter of the SAMP AGNs is smaller (0.15 dex) than that of the total sample (i.e., 0.27 dex) as the measurement errors of the SAMP AGNs are much smaller than that of the literature sample. Note that previous \ha{} lag and error measurements were obtained in various ways. Thus, a uniform analysis of the entire sample is required to better constrain the size-luminosity relation and its scatter.


\section{Discussion}\label{s:discuss}
\subsection{Comparison between the \texorpdfstring{\ha{}}{} and continuum luminosities}\label{ss:lumcomp}
\begin{figure}[!thpb]
\centering
\includegraphics[width=0.45\textwidth]{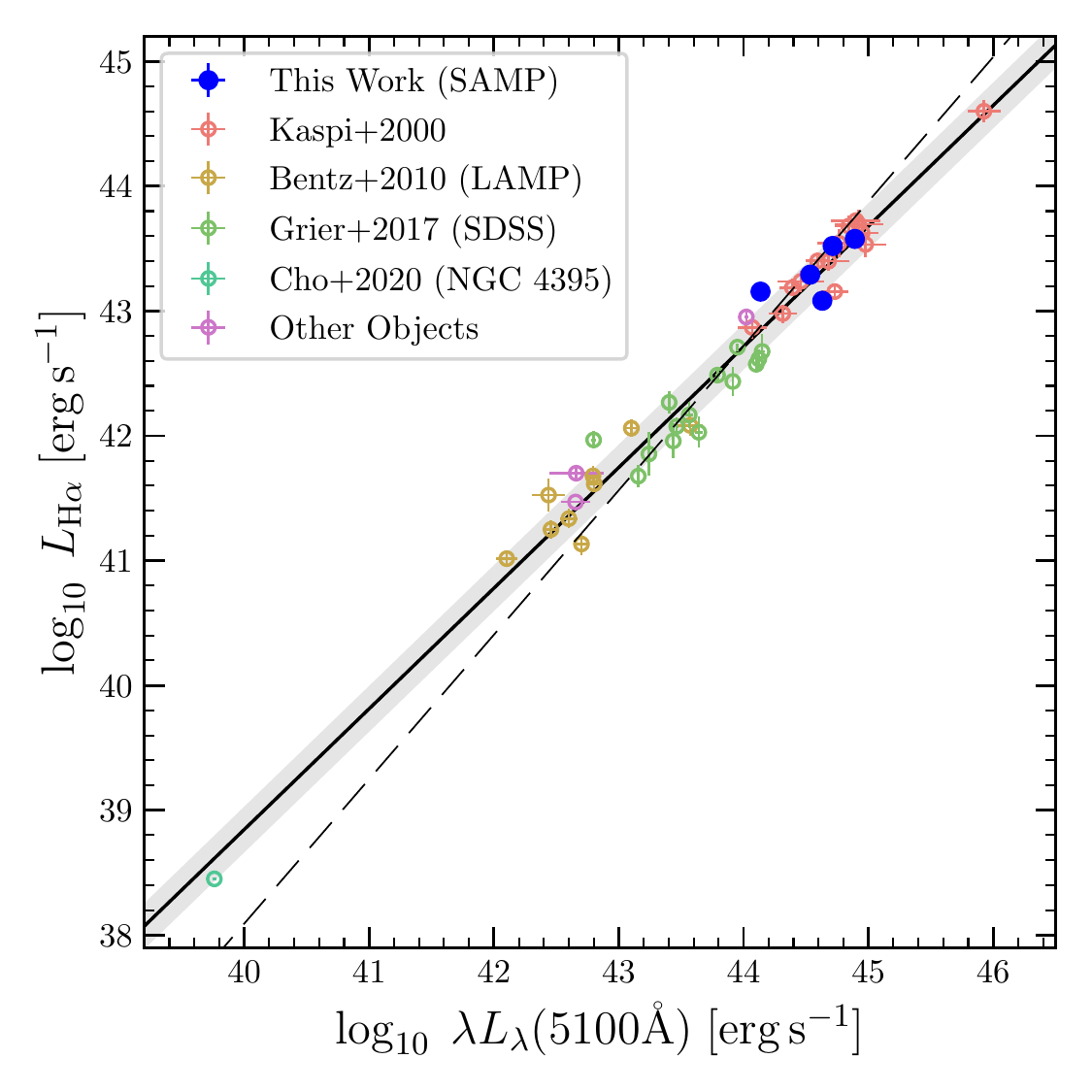}
\caption{The comparison between the continuum luminosity, $\lambda L_\lambda \left({\rm 5100\angstrom{}}\right)$, and the broad \ha{} luminosity, $L_{\rm H\alpha}$. The best-fit relation is denoted with a thick solid line, with the shaded area representing the 1-$\sigma$ intrinsic scatter. The dashed line along the diagonal represents Equation 1 from \citet{Greene&Ho05}.
\label{fig:LhaLcont}}
\end{figure}

The derived size-luminosity relations in \S~\ref{s:sl} have slopes very similar to each other, suggesting that the relation between the \ha{} luminosity and the continuum luminosity at $5100\,\rm\angstrom{}$ is close to a linear one. Also, theoretically, we expect the \ha\ emission line flux to scale linearly with the optical continuum luminosity, assuming that all AGNs share the same spectral slope of the power-law continuum \citep{Yee1980}. We directly compare $L_{\rm H\alpha}$ and $\lambda L_\lambda \left({\rm 5100\angstrom{}}\right)$ in Figure~\ref{fig:LhaLcont} by fitting the relation using the generic relation model given by Eq.~\ref{eq:genericlin}. The best-fit parameters are presented in Table~\ref{table:slfitpar}. The slope $\beta=0.97\pm0.03$ for this relation is virtually the same as unity, and the fit can be simplified as $\lambda L_\lambda \left({\rm 5100 \angstrom{}}\right) = (19\pm 1) L_{\rm H\alpha}$ with an intrinsic scatter of 0.18~dex. The fit parameter did not change even if we excluded NGC~4395 and PG~1226+023 from the fit, suggesting that a simple linear relation is valid across 7 orders of magnitude of luminosities ($10^{39} < \lambda L_\lambda \left({\rm 5100 \angstrom{}}\right) / [{\rm erg\,s^{-1}}] < 10^{46}$, or $10^{38} < L_{\rm H\alpha} / [{\rm erg\,s^{-1}}] < 10^{45}$).

The linearity of this relation contradicts the results by \citet{Greene&Ho05}, who found a supra-linear relation between the continuum and the line luminosities, with a slope of $1.157\pm 0.005$. Note that their sample consists of the AGNs from the Sloan Digital Sky Survey (SDSS), without reverberation mapping results. The major difference compared to our study is that they did not subtract the host galaxy contamination in the continuum luminosity or remove the narrow line components for measuring the broad \ha{} luminosity. In the case of the narrow line contamination, \citet{Greene&Ho05} discussed that the narrow component contribution is only $\sim$7\%{} of the \ha{} flux, and the slope of the relation did not change even if they used the luminosity of the broad \ha{} line only. On the other hand, the host contamination can systematically change the slope. While the continuum subtraction was not performed for some AGNs in Table~\ref{table:fitdata} in our study, these AGNs are mostly high-luminosity objects ($\lambda L_\lambda \left({\rm 5100 \angstrom{}}\right)>10^{44}\,\mathrm{erg\,s^{-1}}$), for which host galaxy contribution would be negligible. However, \citet{Greene&Ho05} did not subtract the host galaxy contribution for their sample over the entire luminosity range, including low-luminosity AGNs where the host galaxy contribution is expected to be systematically larger. While they did exclude galaxies with high host contamination based on modeling the equivalent width of the \ion{Ca}{2} K-line, the model was constructed from the library of early-type galaxies only, which are not expected to be the host galaxies of low-luminosity AGNs. Presumably these are why we obtained a different slope compared to that of \citet{Greene&Ho05}. A more recent study by \citet{Rakshit+20}, which decomposed SDSS spectra to acquire host-free AGN continuum and broad \ha{} line luminosities, also demonstrated a similar supra-linear slope of $1.126\pm 0.004$. However, they decomposed the host spectra using the eigenspectra constructed from late-type galaxies, which can lead to a similar template mismatch. On the other hand, studies that did not use SDSS spectra deduced slopes close to unity. For example, \citet{Shen&Liu12} obtained a slope of $1.010\pm 0.042$ from 60 objects in $z\sim$1.5-2.2. Similarly, \citet{Jun+15} demonstrated a slope of $1.044\pm 0.008$ with AGNs in $z\sim$0-6.2 and continuum luminosity between $10^{42}<\lambda L_\lambda \left({\rm 5100 \angstrom{}}\right)/[{\rm erg\,s^{-1}}]<10^{47}$. We discuss that the slope $\beta$ is unity since (1) all contradicting studies carried out using SDSS spectra had issues with template mismatch, (2) other studies support our results, and (3) the slope of unity predicts the \ha{} luminosity of NGC 4395 well from its continuum luminosity, while the supra-linear slope would underestimate its \ha{} luminosity.

\subsection{Broad line region stratification}\label{ss:stratification}
\begin{figure}[!thpb]
\centering
\includegraphics[width=0.45\textwidth]{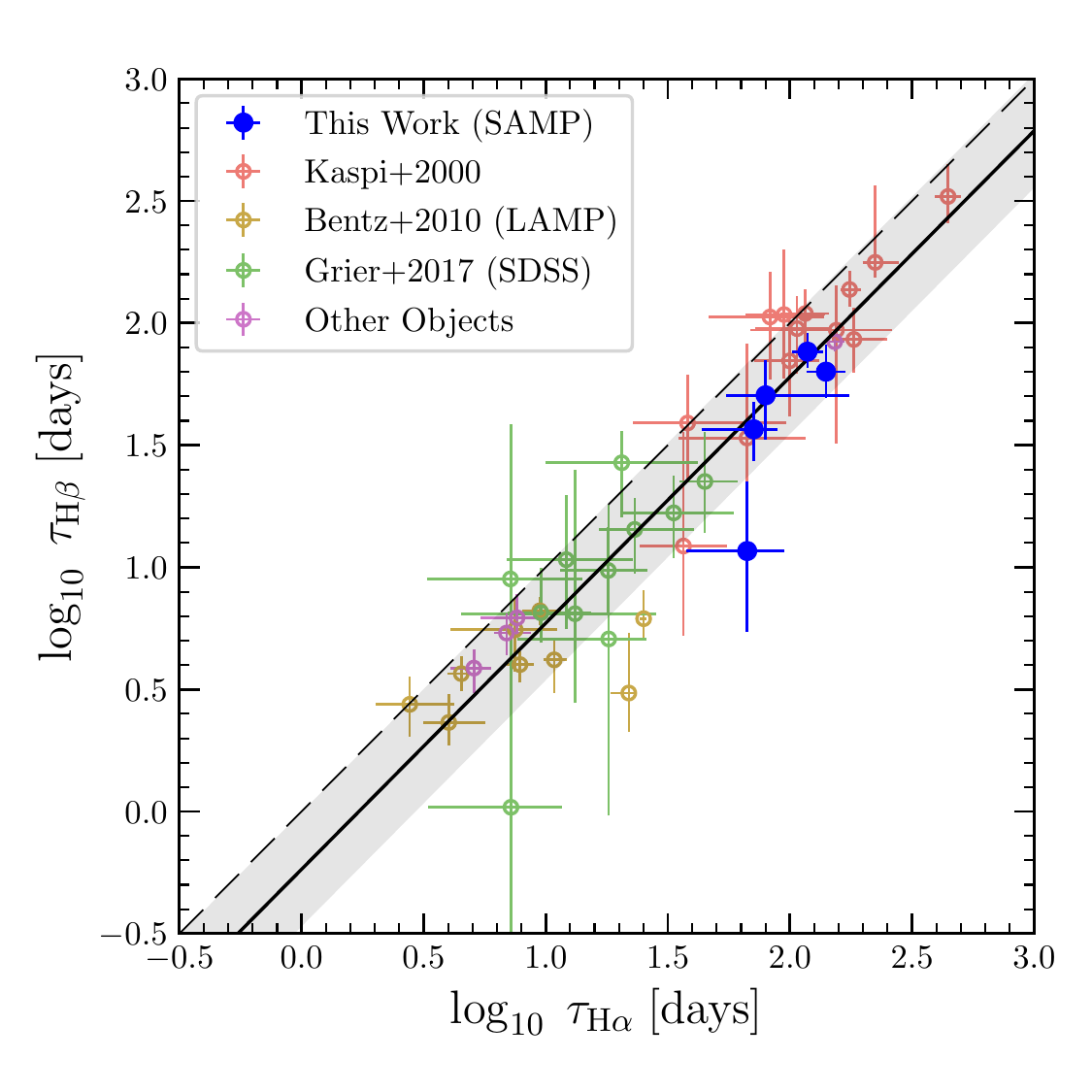}
\caption{The comparison between the \ha{} and \hb{} lags. The best-fit relation is denoted with a thick solid line, with the shaded area representing the 1-$\sigma$ intrinsic scatter. The dashed line along the diagonal represents the $\tau_{\rm H\alpha}=\tau_{\rm H\beta}$ line.\label{fig:hahb}}
\end{figure}

We now compare the time lag of \ha{} with that of \hb{}. For our 5 objects, we adopted the \hb{} time lags in our previous paper \citep{samp_hb_inprep}. Most of the other objects also have \hb{} lags measured as well, which are summarized in Table~\ref{table:fitdata}. The comparison between two lags is presented in Figure~\ref{fig:hahb}. The best-fit parameters for these two lags are listed in Table~\ref{table:slfitpar}, which can be summarized as $\tau_{\rm H\alpha}=1.68 \tau_{\rm H\beta}$ with an intrinsic scatter of $\sigma_{\rm int}=0.23\pm0.03$ dex, where the ratio does not depend on the lag. This shows stratified BLRs across a wide range of time lags, which has been predicted and observed before (e.g., \citealt{Bentz+10} and references therein), and can be attributed to the optical depth of \ha{} being larger than that of \hb{}.

\subsection{Slope of the size-luminosity relation}
The slopes of the \ha{} size-luminosity relation we found in \S~\ref{s:sl}, $0.61\pm 0.04$ and $0.59\pm 0.04$ are consistent with those of the \hb{} size-continuum luminosity relation by \citet[$0.55\pm 0.03$ with \emph{Clean+ExtCorr} sample]{Bentz+13}. This is expected since (1) the \ha{} lags are proportional to the \hb{} time lags, as discussed in \S~\ref{ss:stratification}, and (2) the \ha{} luminosity is proportional to the continuum luminosity, as discussed in \S~\ref{ss:lumcomp}.

While the \ha{} size - \ha{} luminosity relation has not previously been reported, some studies have compared the \hb{} BLR size with \hb{} luminosity. \citet{Wu+04} obtained the slope of $0.684\pm0.106$ based on available \hb{} lags, including a large number of objects from \citet{Kaspi+00}. A similar slope, $0.687\pm0.063$, was obtained by \citet{Kaspi+05}. On the other hand, more recent studies support the \hb{} size-\hb{} luminosity relation having a slope close to that of \citet{Bentz+13}. For instance, \citet{Greene+10} obtained a much smaller slope of $0.53\pm0.04$. \citet{Du+15} also obtained a smaller slope of $0.51\pm0.03$ with a much larger sample. Our result is broadly consistent with these studies, although further studies are needed to determine the slope of the size-luminosity relation using the broad \hb{} luminosity.

\subsection{Single-epoch mass comparison and its implications for IMBH studies}\label{ss:mass}
The mass of the black hole can be estimated by Eq.~\ref{eq:mbh}. We propose a new single-epoch mass estimator using the luminosity and velocity of broad \ha{},
\begin{equation}
\begin{aligned}
\frac{\Mbh}{10^6 M_\odot} &= \left(13\pm 1\right)\\&\times \left(\frac{\sigma_{\rm H\alpha}}{10^3\,\rm km\,s^{-1}}\right)^2\left(\frac{L_{\rm H\alpha}}{10^{42}\,\rm erg\,s^{-1}}\right)^{0.61\pm 0.04} \label{eq:mass_sigma}
\end{aligned}
\end{equation}
\begin{equation}
\begin{aligned}
\frac{\Mbh}{10^6 M_\odot} &= \left(3.2\pm 0.3\right)\\&\times \left(\frac{\rm FWHM_{\rm H\alpha}}{10^3\,\rm km\,s^{-1}}\right)^2\left(\frac{L_{\rm H\alpha}}{10^{42}\,\rm erg\,s^{-1}}\right)^{0.61\pm 0.04} \label{eq:mass_fwhm}
\end{aligned}
\end{equation}
where we adopted $f=4.47$ for $\sigma$ and $f=1.12$ for FWHM from \citet{Woo+15}.

In comparison, \citet{Greene&Ho05} proposed a method to estimate the size of \ha{} as follows. The continuum luminosity is first estimated from \ha{} broad line luminosities using Eq.~1 of their paper. Then, the \hb{} size - continuum luminosity relation is used to estimate the \hb{} BLR size. A commonly adopted relation is Eq.~2 from \citet{Bentz+13} with the parameters obtained using the \emph{Clean+ExtCorr} sample. Combining these yields
\begin{equation}
\begin{aligned}
\log_{10}& \left(\frac{\tau_{\rm H\beta}}{1\,[{\rm day}]}\right) \;\;\;\mbox{(Greene \& Ho + Bentz)}
\\&=1.214 + 0.472 \,\log_{10}\left(\frac{L_{\rm H\alpha}}{10^{42}\,[{\rm erg\,s^{-1}}]}\right)
\end{aligned}
\end{equation}
which will hereafter be denoted as GH+B. Using this along with Eq.~3 by \citet{Greene&Ho05}, one can also construct mass estimates similar to Eqs.~\ref{eq:mass_sigma} and \ref{eq:mass_fwhm}. An example of the analog relation can be found in \citet{Reines+13}.

However, the new mass estimator predicts a substantially different mass from the one the GH+B estimator predicts. In Figure~\ref{fig:slrel}, the GH+B relation is plotted as a dashed line, along with the size-luminosity relation we obtained as a solid line. While the difference between two relations is larger than 0.5 dex if $L_{\rm H\alpha} \leq 10^{38}\,{\rm erg\,s^{-1}}$, the \ha{} size of NGC 4395 is even smaller than what our relation predicts. This difference can be understood as the direct result of the supra-linear relation between $L_{\rm H\alpha}$ and $\lambda L_\lambda \left({\rm 5100 \angstrom{}}\right)$ by \citet{Greene&Ho05}, as discussed in \S~\ref{ss:lumcomp}.

\begin{figure}[!thpb]
\centering
\includegraphics[width=0.45\textwidth]{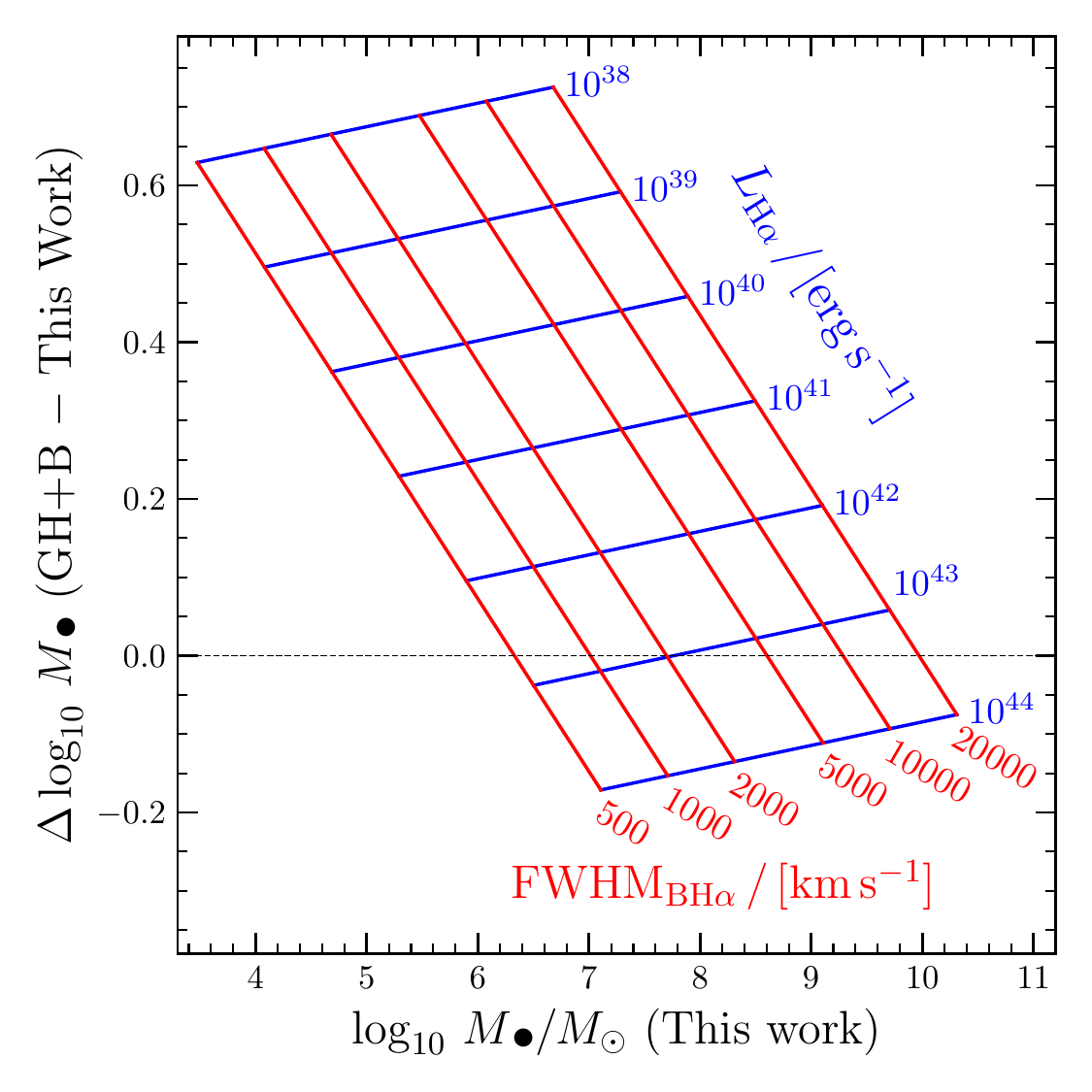}
\caption{Demonstration of \mbh{} overestimation depending on different parameters. The abscissa represents the single-epoch mass of the black hole estimated from \ha{} luminosities using the relation we obtained, while the ordinate represents the excess of the mass when estimated using \citet{Greene&Ho05} and \citet{Bentz+13} compared to our estimate. The red lines show the contours with the same \ha{} line widths, while the blue lines show the contours with the same \ha{} luminosity. \label{fig:mass}}
\end{figure}

To quantify the difference, we compared our mass estimate with that based on the GH+B relation in Figure~\ref{fig:mass}. While it is negligible for AGNs with masses above $10^9\,$\msun{}, the black hole mass can be overestimated by more than a factor of two with the GH+B relation for low-luminosity AGNs with $L_{\rm H\alpha} \leq 10^{40}\,{\rm erg\,s^{-1}}$, which are strong candidates for IMBHs.

IMBHs are important objects because of their relevance to the formation of SMBHs since different scenarios of SMBH formation predict different IMBH mass functions and mass scaling relations even in the local universe \citep[\mbh-$M_\ast$, \mbh-$\sigma_\ast$, etc.; e.g.,][]{Miller+15, Greene+20}. In particular, according to the direct collapse scenario, IMBHs at the centers of galaxies are predicted to have minimum masses of about $10^4$-$10^5$~${\rm M_\odot}$ \citep[e.g.,][]{Ferrara+14}.

If our new mass estimator is accurate, the mass measurements of active IMBHs in the literature may have been biased toward larger values. For example, using the GH+B relation, \citet{Reines+13} estimated the masses of AGNs in dwarf galaxies to be in the range of $10^5\,$\msun{}--$10^6\,$\msun{}. However, considering that the \ha{} luminosities of these AGNs range from $10^{38}\,\rm erg\,s^{-1}$ to $10^{40.5}\,\rm erg\,s^{-1}$, it is possible that the reported masses are overestimated by an average factor of 2-3, with some potentially having a mass closer to $30,000$ \msun{}.

The slope of the \ha{}-based mass estimator is of critical importance. However, our combined sample does not include any object with $10^{39}\,\mathrm{erg\,s^{-1}}<L_{\mathrm{H\alpha}}<10^{41}\,\mathrm{erg\,s^{-1}}$, except for one object (NGC 4395) with a luminosity smaller than this range. It is necessary to include low-luminosity AGNs, which are more representative of active IMBHs, to properly constrain the slope of the size-luminosity relation and the masses of IMBHs.

Conducting a reverberation mapping campaign for AGNs with \ha{} luminosities below $10^{41}\,\mathrm{erg\,s^{-1}}$ poses substantial challenges. First, these low luminosity AGNs are expected to exhibit very short \ha{} lags, requiring intra-night monitoring campaigns with a cadence of several hours or even minutes and continuous observations in a several-day time baseline. Such campaigns require the coordination of multiple telescopes. Second, observing these low luminosity AGNs requires higher sensitivity, which implies the use of larger aperture telescopes and/or longer exposure times. Despite these difficulties, monitoring campaigns for these low luminosity AGNs will be crucial to better constrain the \ha{} S-L relation and to calibrate the \ha{}-based mass estimators, particularly for IMBHs.


\section{Conclusions}\label{s:concl}
We present the \ha{} reverberation mapping results from the Seoul National University AGN Monitoring Project. While the SAMP mainly aims at performing \hb\ reverberation mapping for more than 30 high-luminosity AGNs (Woo et al. in preparation), we additionally obtained time series of \ha{} spectra for 6 objects and performed the reverberation mapping analysis. By combining our new measurements with the \ha{} lag measurements of other AGNs in the literature, we investigated the size-luminosity relation of the broad \ha{} line. Our main results are summarized as follows.
\begin{itemize}
\item We produced \ha{} light curves based on the spectral modeling of \ha{} emission line and measured the time lags against B band continua for 6 new objects, 5 of which we consider to be reliable.
\item We collected a sample of AGNs with the lag and flux measurements of broad \ha{} of 47 AGNs, consisting of our 5 new objects and 42 from the literature. We calculated the \ha{} luminosities after correcting for Galactic extinctions.
\item We found the relation between \ha{} BLR sizes and \ha{} luminosities to be $\log_{10}{\tau_{\rm H\alpha}}/{1\,{\rm day}} = \left(1.16\pm 0.05\right) + \left(0.61\pm 0.04\right)\log_{10}{L_{\rm H\alpha}}/{10^{42}\,{\rm erg\,s^{-1}}}$ and the relation between \ha{} sizes and 5100$\rm\angstrom{}$ continuum luminosities to be $\log_{10}{\tau_{\rm H\alpha}}/{1\,{\rm day}} = \left(1.59\pm 0.05\right) + \left(0.58\pm 0.04\right)\log_{10}{\lambda L_\lambda \left({\rm 5100\angstrom{}}\right)}/{10^{44}\,{\rm erg\,s^{-1}}}$.
\item We found that $\lambda L_\lambda\left({\rm 5100\angstrom{}}\right)=19\,L_{\rm H\alpha}$, and $\tau_{\rm H\alpha}$:$\tau_{\rm H\beta}$=1.68:1.
\item The size-luminosity relation we obtained based on the reverberation mapping results deviates from what is proposed based on single-epoch spectra by \citet{Greene&Ho05}. We demonstrate that for AGNs in the IMBH mass regime, the black hole mass could be substantially overestimated by a factor of 3 on average if the relation by \citet{Greene&Ho05} is used.
\item We propose two mass estimators based on \ha{} broad lines assuming $f=4.47$ for $\sigma$ and $f=1.12$ for FWHM \citep{Woo+15}, \begin{equation*}\begin{aligned}\frac{\Mbh}{10^6 M_\odot} &= \left(13\pm 1\right)\\&\times \left(\frac{\sigma_{\rm H\alpha}}{10^3\,\rm km\,s^{-1}}\right)^2\left(\frac{L_{\rm H\alpha}}{10^{42}\,\rm erg\,s^{-1}}\right)^{0.61\pm 0.04} \end{aligned}\end{equation*}\begin{equation*}\begin{aligned}\frac{\Mbh}{10^6 M_\odot} &= \left(3.2\pm 0.3\right)\\&\times \left(\frac{\rm FWHM_{\rm H\alpha}}{10^3\,\rm km\,s^{-1}}\right)^2\left(\frac{L_{\rm H\alpha}}{10^{42}\,\rm erg\,s^{-1}}\right)^{0.61\pm 0.04} \end{aligned}\end{equation*}
\end{itemize}
\newpage


\begin{acknowledgments}
We thank the anonymous referee for constructive comments that improved the manuscript.
This work has been supported by the Basic Science Research Program through the National Research Foundation of Korean Government (2021R1A2C3008486) and the Samsung Science \& Technology Foundation under Project Number SSTF-BA1501-05.
The work of H.C. was supported by an NRF grant funded by the Korean Government (NRF-2018H1A2A1061365-Global Ph.D. Fellowship Program).
S.W. acknowledges the support from the National Research Foundation of Korea (NRF) grant funded by the Korean government (MEST) (No. 2019R1A6A1A10073437).
V.N.B. gratefully acknowledges assistance from the National Science Foundation (NSF) Research at Undergraduate Institutions (RUI) grant AST-1909297. Note that the findings and conclusions do not necessarily represent the views of the NSF.
Research at UCLA was supported by the National Science Foundation through grant NSF-AST 1907208.
Research at UC Irvine was supported by NSF grant AST-1907290.
V.U. acknowledges funding support from NASA Astrophysics Data Analysis Program (ADAP) grant 80NSSC20K0450. Support for program HST-AR-17063.005 was provided by NASA through a grant from the Space Telescope Science Institute, which is operated by the Associations of Universities for Research in Astronomy, Incorporated, under NASA contract NAS526555.
S.R. acknowledges the partial support of SRG-SERB, DST, New Delhi through grant no. SRG/2021/001334.
We thank the staff of the observatories where data were collected for their assistance.

\end{acknowledgments}

\bibliography{bib.bib}

\end{document}